\documentclass[12pt]{article}
\usepackage[T1]{fontenc}
\usepackage[utf8]{inputenc}
\usepackage[english]{babel}
\usepackage{orcidlink}
\usepackage{bm}
\usepackage{amsmath}
\usepackage{color}
\usepackage{graphicx}
\usepackage{caption}
\usepackage{subfig}
\usepackage{amssymb}
\usepackage{float}
\usepackage{cite}
\usepackage{hyperref}

\graphicspath{{./plots/}} % path for plots

\definecolor{blue}{rgb}{0.27, 0.42, 0.81}
\usepackage{hyperref}
\hypersetup{
colorlinks, linktocpage, bookmarksnumbered,
pdfstartview=FitH,
citecolor=blue,
urlcolor=blue,
linkcolor=red,
filecolor=blue
}

\newcommand{\g}{\gamma}
\newcommand{\GG }{\Gamma_{12}}

\newcommand{\Emiss}{E_\text{miss}}

\begin{document}
\allowdisplaybreaks

\begin{titlepage}

\begin{flushright}
{\small
USC-TH-2024-02\\
\today \\
}
\end{flushright}

\vskip1cm
\begin{center}
{\Large \bf\boldmath Constraints on light dark sector particles from 
\\[0.1cm]  lifetime difference of heavy neutral mesons}
\end{center}

\vspace{0.5cm}
\begin{center}
{ Girish Kumar \orcidlink{0000-0001-6051-2495},
Alexey A. Petrov \orcidlink{0000-0002-4945-4463}} \\[7mm]
{\emph{Department of Physics and Astronomy, \\
University of South Carolina, Columbia, \\
South Carolina 29208, USA}}\\[0.3cm]
\end{center}

\vspace{7mm}
\begin{abstract}
\noindent
Heavy meson decays with missing energy in the final state offer interesting avenues
to search for light invisible new physics such as dark matter (DM). In this context,
we show that such new physics (NP) interactions also affect lifetime difference in neutral
meson-antimeson mixing. We consider general dimension-six effective quark interactions
involving a pair of DM particles and calculate their contributions to lifetime difference
in beauty and charm meson systems. We use the latest data on mixing observables to constrain
the relevant effective operators. We find that lifetime differences
provide novel and complementary flavor constraints compared to those obtained
from heavy meson decays.
\end{abstract}
\end{titlepage}

%\email{girish89@sc.edu}
%%%%%%%%%%%%%%%%%%%%%%%%%%%%%%%%%%%%%%%%%%%%%%%%%%%%%%%%%%%%%%%%%%%
\tableofcontents
%\newpage

%%%%%%%%%%%%%%%%%%%%%%%%%%%%%%%%%%%%%%%%%%
%%%%%%%%%%%%%%%%%%%%%%%%%%%%%%%%%%%%%%%%%%
\section{Introduction}
It is a general consensus at present that a large part of our Universe is composed of
dark matter (DM) \cite{Planck:2015fie,Planck:2018vyg}.
The evidence of DM  comes primarily from the observation of
its gravitational effects, but has been corroborated by several
independent methods \cite{Clowe:2003tk,Rubin:1980zd}.
However, other than its existence and weakly interacting nature,
little is known about its mass and properties, making the detection a challenging endeavor.
There are dedicated experiments searching for DM in distinctive mass windows.
In principle, DM does not have to be composed of a single species of particles,
giving rise to more complicated dark sectors (DS), where one or a few particles 
play the role of DM\footnote{In this paper, we will be using the terms ``dark matter particles'' and ``dark sector particles'' interchangeably, denoting the states that do not interact with the detectors and/or decay to the SM particles inside the detector.}.

If such DS particles are sufficiently light and couple to quarks,
they can also be produced in decays of beauty
and charm mesons \cite{Bird:2004ts,Badin:2010uh}, presenting new intriguing opportunities
for searching DM at low-energy flavor experiments.
These decays are characterized by the missing energy $\Emiss$ in the final state as 
DS particles produced in the decay escape undetected.
In Table \ref{tab: exp-data}, we summarize current data on
relevant charm and beauty mesons~decay~modes.

These decays proceed via flavor-changing neutral current
transitions, which arise only at the loop level in the Standard Model (SM),
with missing energy $\Emiss$ usually corresponding to
a pair of massless neutrinos\footnote{For invisible decay modes $(B^0, B_s^0,  D^0) \to \Emiss$,
the SM contribution is dominated by $4\nu$ final state\cite{Bhattacharya:2018msv}.}.
The SM predictions for these processes are quite small,
as can be seen from Table \ref{tab: exp-data}.
Furthermore, as neutrinos are present in the final state,
the corresponding theoretical predictions
do not suffer from uncertainties arising due to photon exchange.
Precise measurements of these decay modes, therefore, allow for highly sensitive
probe of new physics (NP) beyond the SM, including light DM. 

Several papers have focused on these decay modes as probes of light invisible NP
(see for example,
Refs.~\cite{
Bird:2004ts,Altmannshofer:2009ma,Kamenik:2011vy,Tandean:2019tkm,Li:2020dpc,
Fajfer:2021woc,Li:2021sqe,Felkl:2021uxi,He:2022ljo,Geng:2022kmf,Li:2023sjf,
Felkl:2023ayn,He:2023bnk,Gabrielli:2024wys,Bolton:2024egx,Buras:2024ewl}). 
Here, it is worth mentioning that very recently
Belle-II found the evidence for
$B^+ \to K^+ \Emiss$, measuring the branching ratio $(2.3\pm 0.7) \times 10^{-5}$.
The value is higher than the corresponding SM value at a level $2.7\sigma$, and
can be explained if there are light invisible NP contributions to the decay,
where, for example, $\Emiss$ is carried away by DM, in addition to SM neutrinos.
In this light, there has been reinvigorated interest and motivation
to study decays listed in Table \ref{tab: exp-data}.

%================================================
{\renewcommand{\arraystretch}{1.02}
\begin{table}[t]
\begin{center}
\begin{tabular}{l@{\hspace{3em}}c@{\hspace{3em}}r}
\hline
\hline
Decay mode  & SM prediction & Current bound  \\	
\hline
$D^0 \to \Emiss$ & $\sim 10^{-27}$ \cite{Bhattacharya:2018msv} & $ 9.4 \times 10^{-5}$ \quad (Belle \cite{Belle:2016qek})   \\
$D^0 \to \pi^0 \Emiss$ & $\sim 10^{-16}$ \cite{Burdman:2001tf} & $2.1 \times 10^{-4}$  (BESIII \cite{BESIII:2021slf})   \\
\hline
$B_s^0 \to  \Emiss$ & $\sim 5 \times 10^{-15}$ \cite{Bhattacharya:2018msv}& $5.6 \times 10^{-4}$ \cite{Alonso-Alvarez:2023mgc} \\
$B^+ \to  K^+ \Emiss$ & $(4.71 \pm 0.24) \times 10^{-6}$ &$(2.3 \pm 0.7) \times 10^{-5}$ (Belle-II \cite{Belle-II:2023esi}) \\
$B^0 \to  K^0 \Emiss$ &$(4.35 \pm 0.21) \times 10^{-6}$ & $2.6 \times 10^{-5}$ \quad(Belle \cite{Belle:2017oht}) \\
$B^0 \to  K^{\ast 0} \Emiss$ &$(9.81 \pm 0.95) \times 10^{-6}$ &$1.8 \times 10^{-5}$\quad(Belle \cite{Belle:2017oht}) \\
$B^+ \to  K^{\ast +} \Emiss$ & $(1.06 \pm 0.10) \times 10^{-5}$ &$4.0 \times 10^{-5}$ \quad(Belle \cite{Belle:2013tnz}) \\
\hline
$B^0 \to  \Emiss$ & $\sim 10^{-16}$ \cite{Bhattacharya:2018msv}& $2.4 \times 10^{-5}$ ~(BaBar \cite{BaBar:2012yut}) \\
$B^0 \to  \pi^0 \Emiss$ & $(6.52 \pm 0.78) \times 10^{-8}$ &$0.9 \times 10^{-5}$ \quad(Belle \cite{Belle:2017oht}) \\
$B^+ \to  \pi^+ \Emiss$ & $(1.40 \pm 0.16) \times 10^{-7}$ &$1.4 \times 10^{-5}$ \quad(Belle \cite{Belle:2017oht}) \\
$B^0 \to  \rho^0 \Emiss$ & $(1.88  \pm 0.35) \times 10^{-7}$& $4.0 \times 10^{-5}$ \quad(Belle \cite{Belle:2017oht}) \\
$B^+ \to  \rho^+ \Emiss$ & $(4.06 \pm 0.73) \times 10^{-7}$& $3.0 \times 10^{-5}$ \quad(Belle \cite{Belle:2017oht}) \\
\hline
\hline
\end{tabular}
\caption{ Summary of experimental data on the branching ratios of charm and beauty mesons
considered in our analysis. All upper bounds are  at 90\% C.L.
The SM predictions, unless noted otherwise, are obtained
using \texttt{Flavio} package \cite{Straub:2018kue}.
Note that while $B^0 \to \Emiss$ and $D^0 \to \Emiss$ are bounded from the 
direct $B$-factory factory measurements, the upper limit on $B_s^0 \to  \Emiss$ in
Ref.~\cite{Alonso-Alvarez:2023mgc} is obtained indirectly using ALEPH data \cite{ALEPH:2000vvi}.
}
\end{center}
\end{table}\label{tab: exp-data}
}
%================================================ 

The purpose of this article is to point out that quark-DM
interactions relevant to these decays also \emph{unavoidably}
generate a finite NP contribution to the lifetime difference in
meson-antimeson mixing, which provides additional constraints on light DM.
In particular, we show that lifetime difference can be used to obtain novel bounds on
certain quark-DM interactions that remain unconstrained so far.
In order to see how the contribution to lifetime difference arises from above-mentioned
quark-DM interactions, let us consider the off-diagonal term
of the neutral meson $M^0$  mass matrix \cite{Golowich:2006gq,Golowich:2007ka}
\begin{align}\label{eq: mix12}
	\left(M - \frac{i}{2}\Gamma\right)_{12} = \frac{1}{2 m_{M^0}}\left[
	\langle {M^0} |\mathcal{H}_\text{eff}^{|\Delta F|=2}| \overline{M^0}\rangle
	+ \sum_n\frac{\langle {M^0} |\mathcal{H}_\text{eff}^{|\Delta F|=1}|n\rangle\langle n |\mathcal{H}_\text{eff}^{|\Delta F|=1}| \overline{M^0}\rangle}{m_{M^0} - E_n + i \epsilon}\right],
\end{align}
where $M_{12}$ denotes the dispersive part of mixing amplitude,
which contributes to the $M^0$-$\overline{M^0}$ mass difference,
whereas $\Gamma_{12}$ is related to absorptive part
contributing to lifetime difference in $M^0$ and $\overline{M^0}$,
with $M^0= B^0, B_s^0$, or $D^0$ mesons. 
On the right hand side, the first term concerns local $|\Delta F|=2$ interactions and
does not have absorptive part. However, the second term, which involves 
intermediate  states $n$ and concerns $|\Delta F|=1$ transitions,  generates
an absorptive piece, which shows that the lifetime difference is governed by
$|\Delta F|=1$ interactions. 

The paper is organized as follows. In the next section
we define general effective Hamiltonian containing all relevant quark-DM interactions. 
In Section \ref{sec: lifetime-diff} we calculate DM contribution to the lifetime difference
in heavy meson-mixing.
In Section \ref{sec: results}, we present our results: we constrain the 
quark-DM interactions
using current data on lifetime differences, and compare these with those from
$c\to u\Emiss$, $b\to \{s, d\}\,\Emiss$ decays.
Finally, we summarize our conclusions in Section \ref{sec: summary}.

%%%%%%%%%%%%%%%%%%%%%%%%%%%%%%%%%%%%%%%%%%
%%%%%%%%%%%%%%%%%%%%%%%%%%%%%%%%%%%%%%%%%%
\section{Relevant effective interactions}\label{sec: effective-interactions}

To keep our analysis general, we  work in the framework of effective field 
theory (EFT) valid
at some appropriate low energy scale that is typically set by the mass of decaying meson.
Note that the final state in the decay can, in principle, contain a single or more DM 
or DS fields. The underlying quark-DM interactions will also have one or more 
DM fields, accordingly.
However, operators with a single DM field can only generate an absorptive part 
of the meson mixing amplitude if they have DM mass close to that of $B_s^0$, $B^0$, 
or $D^0$ states and a large width \cite{Golowich:1998pz}. Since, by definition, 
DM particles should have cosmological-scale lifetimes, we will not consider such 
operators here.
Furthermore, we restrict the scope of our analysis to effective operators of 
dimension up to six. The general effective Hamiltonian can then be written as
\begin{align}\label{eq: Heff}
	\mathcal{H}_\text{eff} =
			\sum_i \frac{C_i (\mu)}{\Lambda^2} \mathcal{O}_i\,,
\end{align}
where $C_i (\mu)$ are the Wilson coefficients of effective operators $\mathcal{O}_i$ 
evaluated at low energy scale $\mu$. The scale $\Lambda$ denotes NP scale, 
which in a specific UV model is typically related to the mass of heavy NP 
particle(s) mediating interactions between the SM and DM fields.
The number of independent operators in Eq.~\eqref{eq: Heff} depends on the spin 
of DM particle.
The full basis of DM operators have been specified previously also (for example,
see Refs.~\cite{Badin:2010uh,Lehmann:2020lcv,He:2022ljo}).
In this paper, we will focus on scenarios where DM is either a scalar or fermion particle.
We follow the operator basis provided in Ref.~\cite{He:2022ljo}, with our notation
similar to Ref.\cite{Lehmann:2020lcv}. Note that our restriction to the operators of
dimension six excludes operators with vector dark sector particles such as 
$\bar q_i Q_j G^{\mu\nu}_D G_{D \mu\nu}$, where $G^{\mu\nu}_D$ is a strength tensor 
of the vector DS field $V^\mu_D$. Note that this restriction, in principle, does not 
exclude the decays of the heavy quark states into two vector DS particles, 
$M^0 \to V_D V_D$.

In case of scalar DM (denoted as $\phi$), there are four operators,
\begin{equation}\label{eq: scalar-ops}
	\begin{aligned}
		\mathcal{O}_{S}^{q_iQ_j} &= m_{Q_j} (\bar q_i Q_j)(\phi^\dagger \phi),
	&\mathcal{O}_{P}^{q_iQ_j}& = m_{Q_j} (\bar q_i i\g_5 Q_j)(\phi^\dagger \phi),\\
	\mathcal{O}_{V}^{q_iQ_j} &= (\bar q_i \g^\mu Q_j)(\phi^\dagger i\overleftrightarrow{\partial_\mu}\phi),
	&\mathcal{O}_{A}^{q_iQ_j}& = (\bar q_i \g^\mu\g_5 Q_j)(\phi^\dagger i\overleftrightarrow{\partial_\mu}\phi),
	\end{aligned}
\end{equation}
where quark indices $\{q_iQ_j\} = \{uc\}$ for $c\to u$, $\{sb\}$ for $b\to s$, and
$\{db\}$ for $b\to d$ quark transitions, and
$\phi^\dagger\overleftrightarrow{\partial_\mu}\phi
= \phi^\dagger(\partial_\mu\phi) - (\partial_\mu\phi^\dagger) \phi$.
Compared to Ref.~\cite{He:2022ljo}, we have multiplied operators $\mathcal{O}_{S, P}^{q_iQ_j}$
by quark mass ($m_{Q_j}$) so that all operators have same dimensions. Note that operators
$\mathcal{O}_{V, A}^{q_iQ_j}$ vanish if $\phi$ is a real field.

In case of the Dirac-type fermionic DM (denoted as $\psi$), the number of effective 
operators of dimension six is larger. These operators are given by
\begin{equation}\label{eq: fermion-ops}
	\begin{aligned}
		\mathcal{O}_{SS}^{q_iQ_j} &= (\overline q_i Q_j)(\bar \psi\psi ), 
	&\mathcal{O}_{PS}^{q_iQ_j}& = (\overline q_i i\g_5 Q_j)(\bar \psi\psi),\\
	\mathcal{O}_{SP}^{q_iQ_j} &= (\overline q_i Q_j)(\bar \psi i\g_5 \psi),
	&\mathcal{O}_{PP}^{q_iQ_j}& = (\overline q_i \g_5 Q_j)(\bar \psi \g_5 \psi),\\
	\mathcal{O}_{VV}^{q_iQ_j} &= (\overline q_i \g^\mu Q_j)(\bar \psi \g_\mu \psi),
	&\mathcal{O}_{AV}^{q_iQ_j}& = (\overline q_i\g^\mu \g_5 Q_j)(\bar\psi\g_\mu \psi),\\
	\mathcal{O}_{VA}^{q_iQ_j} &= (\overline q_i \g^\mu Q_j)(\bar \psi \g_\mu \g_5 \psi),
	&\mathcal{O}_{AA}^{q_iQ_j}& =(\overline q_i\g^\mu\g_5 Q_j)(\bar\psi\g_\mu\g_5 \psi),\\
	\mathcal{O}_{TT}^{q_iQ_j} &= (\overline q_i \sigma^{\mu\nu} Q_j)(\bar\psi \sigma_{\mu\nu} \psi),
	&\mathcal{O}_{T\tilde T}^{q_iQ_j}& = (\overline q_i\sigma^{\mu\nu} Q_j)(\bar\psi\sigma_{\mu\nu}\g_5\psi),
	\end{aligned}
\end{equation}
where $\sigma_{\mu\nu} = i [\g_\mu, \g_\nu]/2$. For the Majorana DM, 
vector current operators $\mathcal{O}_{VV, AV}$ and both tensor current operators vanish.

%%%%%%%%%%%%%%%%%%%%%%%%%%%%%%%%%%%%%%%%%%
%%%%%%%%%%%%%%%%%%%%%%%%%%%%%%%%%%%%%%%%%%
\section{DM contribution to lifetime difference}\label{sec: lifetime-diff}

\begin{figure}[t]
\centering
     \includegraphics[width=6.8cm]{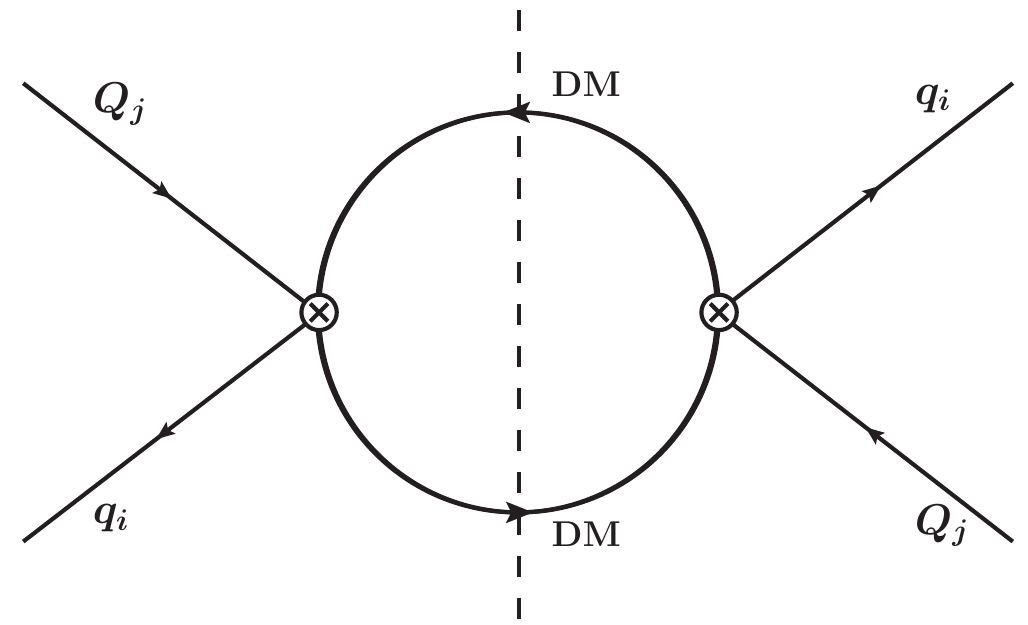}
\caption{Leading order Feynman diagram contributing to $\GG$.}
\label{fig: G12}
\end{figure}

As pointed out in the Introduction, the imaginary part of bi-local contributions in
Eq.~\eqref{eq: mix12} arising due to $|\Delta F|=1$ interactions gives contribution to
the off-diagonal element $\GG$ of decay width matrix in meson-antimeson mixing. 
The optical theorem relates the matrix element $\GG$ to the imaginary part of the 
forward scattering amplitude as
\begin{align}\label{eq: G12}
	\Gamma_{12} = \frac{1}{2\,m_{M^0}}\langle {M^0} | \operatorname{Im} \mathcal{T}| \overline{M^0}\rangle.
\end{align}
Here, $\mathcal{T}$ is the transition operator defined as
\begin{align}
	\mathcal{T} =
	i \int d^4x \, T\left[\mathcal{H}_\text{eff}^{|\Delta F|=1}(x)
	\mathcal{H}_\text{eff}^{|\Delta F|=1}(0)\right],
\end{align}
where $T$ is the time-ordering operator and $|\Delta F|=1$
effective Hamiltonian is given in Eq.~\eqref{eq: Heff}.
The time-ordered product in Eq.~\eqref{eq: G12} is a nonlocal quantity,
which in the heavy quark limit can be expanded in powers of $1/m_Q$ as a series
of local operators. The leading contribution of this operator product expansion
corresponds to the diagram shown in Fig.~\ref{fig: G12}, where each vertex corresponds to
insertion of effective operators given in Eqs.~\eqref{eq: scalar-ops} 
and \eqref{eq: fermion-ops}
for scalar and fermion DM, respectively. Note that since intermediate states are
DM particles, there are no contributions from their interference with the SM operators.

Let us first consider contributions of scalar DM operators in
Eq.~\eqref{eq: scalar-ops}.
The discontinuity of the relevant diagram can be calculated using
Cutkosky rules \cite{Cutkosky:1960sp}.
We utilize  \texttt{Package-X} \cite{Patel:2016fam}
to this end and  to further simplify the
Dirac structures of results.
For the operators $\mathcal{O}_{S}^{q_iQ_j}$
and $\mathcal{O}_{V}^{q_iQ_j}$, we obtain
\begin{equation}\label{eq: G12-scalar}
	\begin{aligned}
			(\GG^{M^0})_{S} &=
		\frac{(C^{q_iQ_j}_{S})^2\, x_{Q_j} \beta_\phi\,m_{M^0}}{32\pi}\,
		\langle Q^{q_iQ_j}_S\rangle, \\
	(\GG^{M^0})_{V} &=
		-\frac{(C^{q_iQ_j}_{V})^2\, \beta_\phi^3\,m_{M^0}}{96\pi}\,
		(  x_{Q_j}\langle  Q^{q_iQ_j}_S\rangle
		+   \langle  Q^{q_iQ_j}_V\rangle),
	\end{aligned}
\end{equation}
respectively, while the corresponding expressions
due to $\mathcal{O}_{P}^{q_iQ_j}$ and $\mathcal{O}_{A}^{q_iQ_j}$ can be
obtained as $(\GG^{M^0})_{P}\to  -(\GG^{M^0})_{S}$ and 
$(\GG^{M^0})_{A}\to  (\GG^{M^0})_{V}$,
respectively, after replacing labels $S\to P$ and $V \to A$ everywhere therein.
Further, $x_{X}  = m_X^2/m_{M^0}^2$,
$\beta_X = \sqrt{1- 4\, x_X}$, and $M^0$
denotes  $B_s^0$, $B^0$, and $\overline{D^0}$ for $\{q_iQ_j\} = \{sb\}$, $\{db\}$,
and $\{uc\}$, respectively.

The quantities $\langle Q^{q_iQ_j}_a\rangle = \langle {M^0} |
(\bar q_i \Gamma^{a} Q_j)(\bar q_i \Gamma_{a}  Q_j)| \overline{M^0}\rangle$
with $\Gamma_a = 1, \g_5, \g_\mu,  \g_\mu\g_5, \sigma_{\mu\nu} $
for $a=S, P, V, A, T$, respectively, are the matrix elements of
$\Delta F=2$ effective operators, and have been defined in the Appendix \ref{app: DeltaF2}.
In our calculation, we take incoming momentum-squared
in the loop $s = (p_{Q_j} + p_{q_i})^2 \simeq m_{M^0}^2$. In addition, we 
also ignore the masses of $s$ and $u$ quarks.

For the fermionic DM operators in Eq.~\eqref{eq: fermion-ops}, we obtain
\begin{align}\label{eq: G12-fermion}
			(\GG^{M^0})_{SS} &=
		\frac{(C^{q_iQ_j}_{SS})^2\,\beta_\psi^3\, m_{M^0}}{16\pi}
		\langle Q^{q_iQ_j}_S\rangle, \nonumber\\
	(\GG^{M^0})_{SP} &=
		\frac{(C^{q_iQ_j}_{SP})^2\,\beta_\psi\, m_{M^0}}{16\pi}
		\langle Q^{q_iQ_j}_S\rangle, \nonumber\\
	(\GG^{M^0})_{VV} &=
		-\frac{(C^{q_iQ_j}_{VV})^2\, (1 + 2 x_\psi)\beta_\psi\,m_{M^0}}{24\pi}
		(  x_{Q_j}\langle Q^{q_iQ_j}_S\rangle 
		+  \langle Q^{q_iQ_j}_V\rangle),\\
	(\GG^{M^0})_{VA} &=
		-\frac{(C^{q_iQ_j}_{VA})^2\, \beta_\psi\,m_{M^0} }{24\pi}
		(  (1 + 2 x_\psi) x_{Q_j}  \langle Q^{q_iQ_j}_S\rangle 
		+   \beta_\psi^2\langle Q^{q_iQ_j}_V\rangle),\nonumber\\
	(\GG^{M^0})_{TT} &=
		-\frac{(C^{q_iQ_j}_{TT})^2\, \beta_\psi\,m_{M^0}}{24\pi}
		\{4(1 + 2 x_\psi) (\langle Q^{q_iQ_j}_S\rangle
		+   x_{Q_j} \langle Q^{q_iQ_j}_V \rangle)  
		-   \beta_\psi^2\langle Q^{q_iQ_j}_T\rangle \},\nonumber\\
	(\GG^{M^0})_{T\tilde T} &=
		\frac{(C^{q_iQ_j}_{T\tilde T})^2\, \beta_\psi m_{M^0}}{24\pi}
		\{4(1 + 2 x_\psi)(\langle Q^{q_iQ_j}_S\rangle 
		+  x_{Q_j} \langle Q^{q_iQ_j}_V \rangle)
		-(1+8x_\psi)\langle Q^{q_iQ_j}_T \rangle\}.\nonumber
\end{align}

The corresponding contributions of remaining operators are obtained as
$(\GG^{M^0})_{PS,\, PP}\to  -(\GG^{M^0})_{SS,\, SP}$,
$(\GG^{M^0})_{AV,\, AA} \to (\GG^{M^0})_{VV,\, VA}$,
after making obvious change of Wilson coefficients and replacing labels $S\to P$ and $V \to A$
in the  matrix elements $\langle Q^{q_iQ_j}_a\rangle$.
The results for $\GG$ in Eqs.~\eqref{eq: G12-scalar} and \eqref{eq: G12-fermion} are general
and applicable to all three heavy meson-antimeson mixing systems: $B_s^0$-$\overline{B_s^0}$,
$B^0$-$\overline{B^0}$, and $D^0$-$\overline{D^0}$.

\textbf{\boldmath $B_s^0$-$\overline{B_s^0}$ and $B^0$-$\overline{B^0}$ systems:}
For neutral $B$ systems, the total width difference $\Delta\Gamma_q$ (with $q=s, d$
for $B_s^0$ and $B^0$ systems, respectively) in presence of NP is
given by \cite{Artuso:2015swg}
\begin{align}
	\Delta \Gamma_{q} = 2 \lvert \GG^{q,\,\rm SM} + \GG^{q,\,\rm NP}\rvert  \cos\phi^q_{12}\,,
\end{align}
with the mixing phase $\phi^q_{12} = \operatorname{arg}(-M_{12}^q/\GG^{q})$. 
In our case, the NP part $\GG^{q,\,\rm NP}$ corresponds to the DM contribution, and for
neutral $B$ systems is parametrized by $(\GG^{M})_i$ in Eqs.~\eqref{eq: G12-scalar}
and \eqref{eq: G12-fermion}. 
On the other hand, to account for the SM part,
we employ the following expression (for a review see Ref.~\cite{Artuso:2015swg})
\begin{align}\label{eq: G12-SM}
	\frac{\GG^q}{M_{12}^q} = 10^{-4}\left[ c_q + a_q \frac{\lambda_u}{\lambda_t} + b_q \left(\frac{\lambda_u}{\lambda_t}\right)^2\right],
\end{align}
with $\lambda_{i} = V_{i b} V_{i q}^\ast$\,, where $V$ is the CKM matrix.
The numerical values of real coefficients $a_q, b_q, c_q$ (notation originally introduced in Ref.~\cite{Beneke:2003az}) are taken from Ref.~\cite{Artuso:2015swg}.
These coefficients exhibit hierarchy $|c_q| \gg |a_q| \gg |b_q|$ 
which together with the smallness of CKM ratio
$|\lambda_u/\lambda_t|\sim 10^{-2}$ indicate that the SM value of
$\GG^q/M_{12}^q$ is dominated by the $c_q$ term. The dispersive part $M_{12}^q$
in the SM is dominated by the short-distance contribution arising from the
box diagram involving top quark and $W$ boson in the loop and given by (see Ref.~\cite{Albrecht:2024oyn} for the latest review)
\begin{align}
	M_{12}^q = \frac{G_F^2}{12\pi^2}\lambda_{t}^2 m_W^2 S_{0}(m_t^2/m_W^2)\hat\eta_B m_{B_q^0} f_{B_q}^2B^{(1)}_{B_q},
\end{align}
where $G_F$ denotes the Fermi constant \cite{ParticleDataGroup:2022pth},
$m_W$ is the mass of $W$ boson, and  Inami-Lim function $S_0(x)$ \cite{Inami:1980fz}
parametrizes top quark loop contribution. Note that top mass $m_t$ here
is in $\overline{MS}$-mass scheme. The factor $\hat\eta_B\simeq 0.84$ \cite{Buras:1990fn}
encodes the renormalization group running from heavy scale $(m_t)$ to $B$ meson mass scale,
whereas $f_{B_q}$
is the $B$ meson decay constant and $B^{(1)}_{B_q}$ is the bag parameter
related to matrix elements of the $\Delta B=2$ SM effective operator
as defined in Appendix \ref{app: DeltaF2}. 

On the other hand, the current measurements of  $\Delta\Gamma_q$ are \cite{HFLAV:2022esi}
\begin{align}
	(\Delta \Gamma_s)_\text{exp} = (0.083 \pm  0.005) ~\text{ps}^{-1},\quad (\Delta \Gamma_d/\Gamma_{d})_\text{exp} = 0.001 \pm  0.010,
\end{align}
where $\Gamma_d$ is the total decay width of $B^0$ meson.

\textbf{\boldmath $D^0$-$\overline{D^0}$ system:}
The width difference in charm system can be
parametrized as $y_{12}^D = |\GG^D|/\Gamma_D$,
%%
%\begin{align}
%	y_{12}^D = \frac{|\GG^D|}{\Gamma_D}\,,
%\end{align}
%%
with $\Gamma_D$ denoting total decay width of $D^0$ meson.
The SM prediction of mass and width difference in neutral charm system
is very tiny due to enhanced CKM suppression and more pronounced GIM cancellations between
internal light quarks (for a recent review, see Ref.~\cite{Lenz:2020awd}).
This allows for charm mixing to be particularly useful for probing
indirect NP effects~\cite{Golowich:2006gq,Golowich:2007ka}. 
On the experimental side, the latest fit results from HFLAV \cite{HFLAV:2022esi} gives
\begin{align}\label{eq: G12-D-exp}
	(y_{12}^D)_\text{exp} = 0.641^{+0.024}_{-0.023} \,\%.
\end{align}

In order to obtain conservative bounds, given the highly suppressed $\GG^D$ in the SM, in our analysis
we will assume that experimental value of $y_{12}^D$ is saturated by the DM contributions.

%%%%%%%%%%%%%%%%%%%%%%%%%%%%%%%%%%%%%%%%%%
%%%%%%%%%%%%%%%%%%%%%%%%%%%%%%%%%%%%%%%%%%
\section{Results}\label{sec: results}

We are now in position to present constraints on NP effective interactions
in Eqs.~\eqref{eq: scalar-ops} and \eqref{eq: fermion-ops}
from lifetime differences in neutral beauty and charm mesons,
and confront them with constraints from branching ratios of decay modes listed
in Table \ref{tab: exp-data}. The details of calculation of decay rate of
meson decays and employed hadronic form factors  have been
relegated to appendix \ref{app: decay-rates}. To keep our analysis simple,
we will work in single operator dominance scenario.  

\subsection{Constraints on scalar DM}\label{subsec: results-scalar}
\begin{figure}[t!]
\centering
     \includegraphics[width=4.7cm]{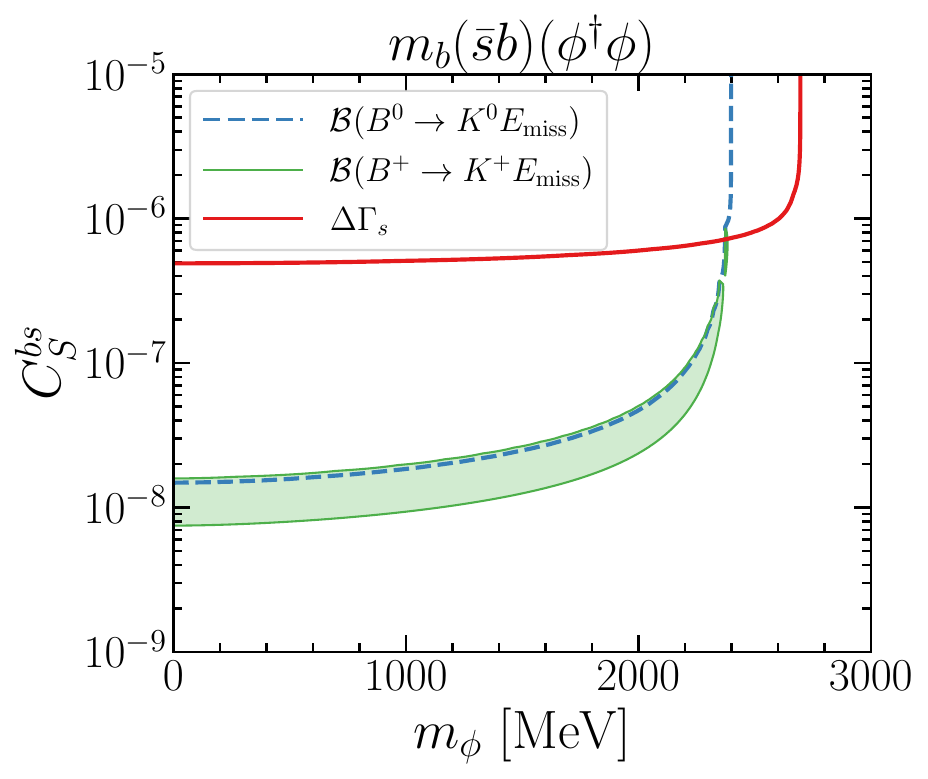}\includegraphics[width=4.7cm]{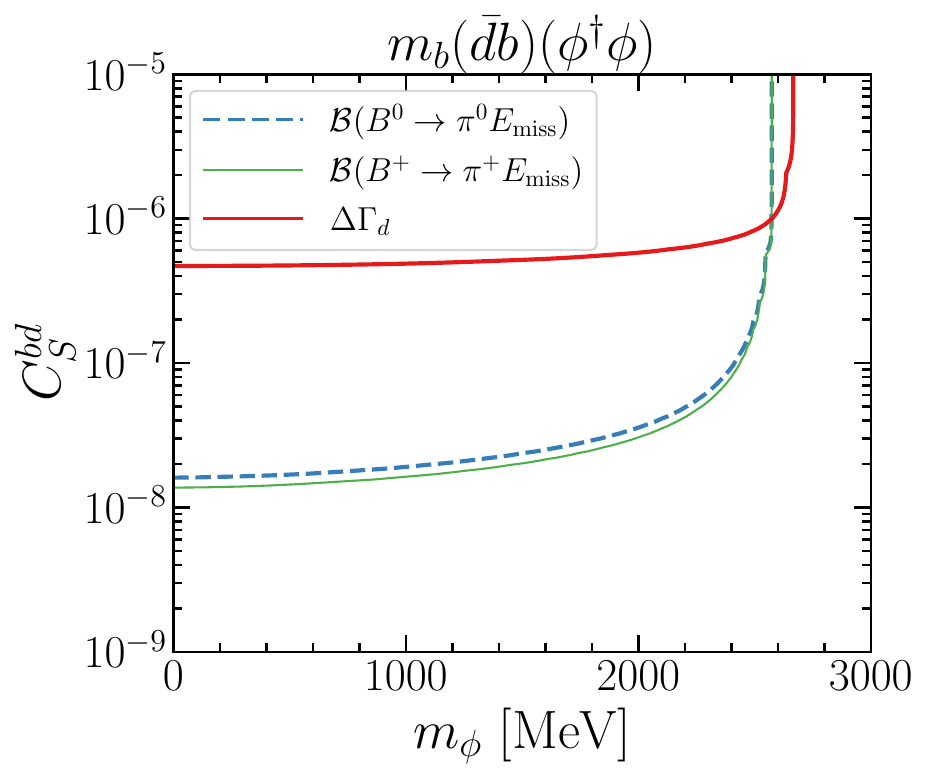}\includegraphics[width=4.7cm]{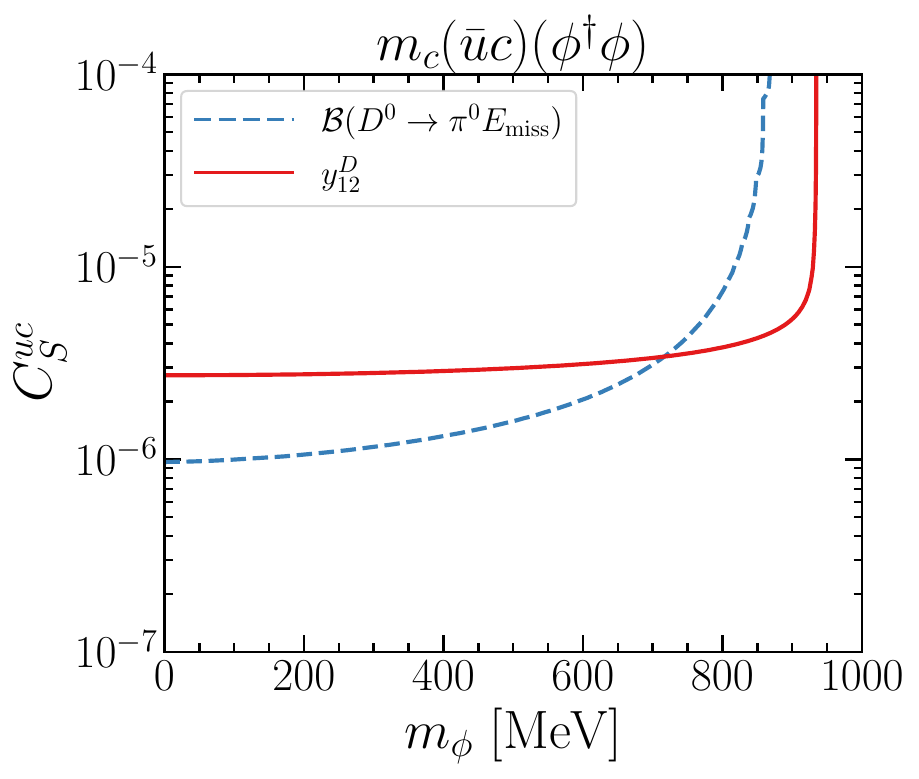}\\
     \includegraphics[width=4.7cm]{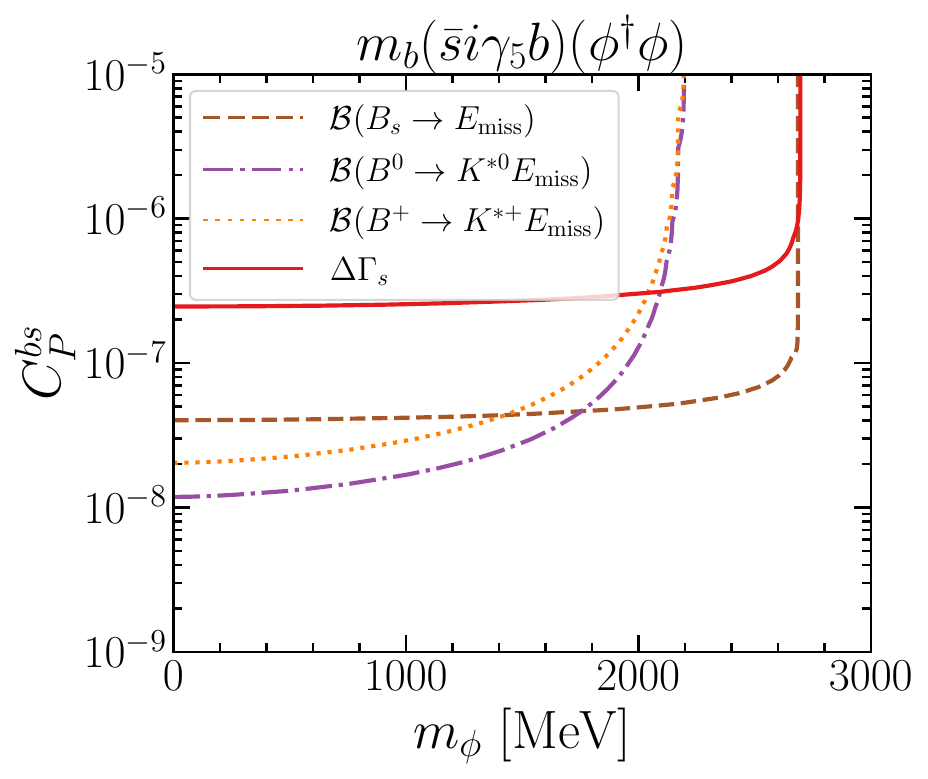}\includegraphics[width=4.7cm]{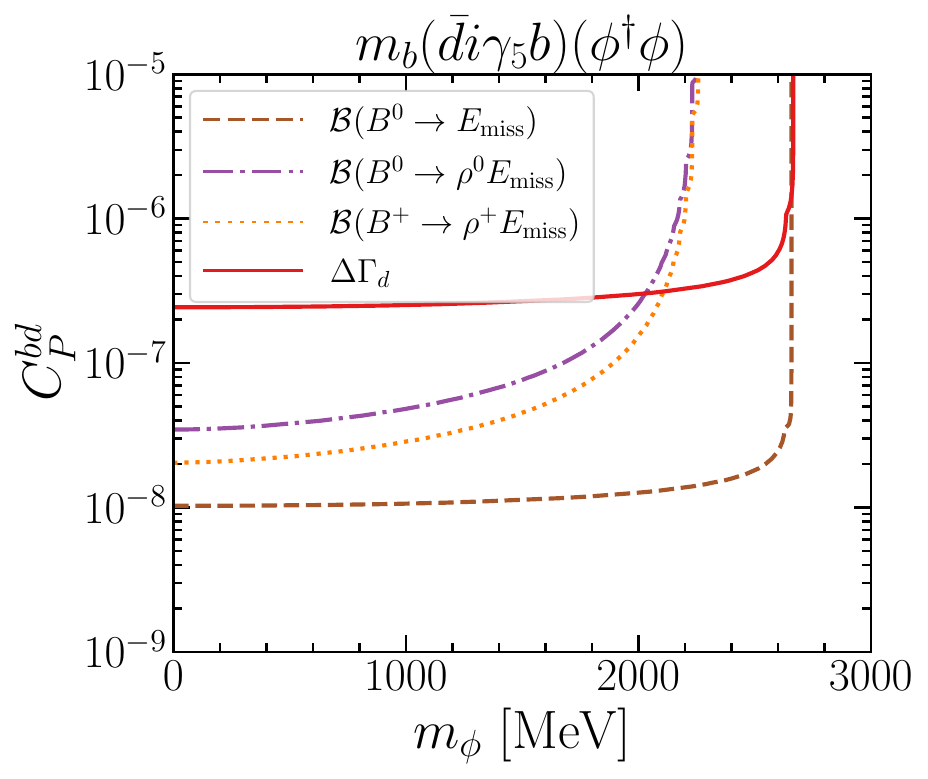}\includegraphics[width=4.7cm]{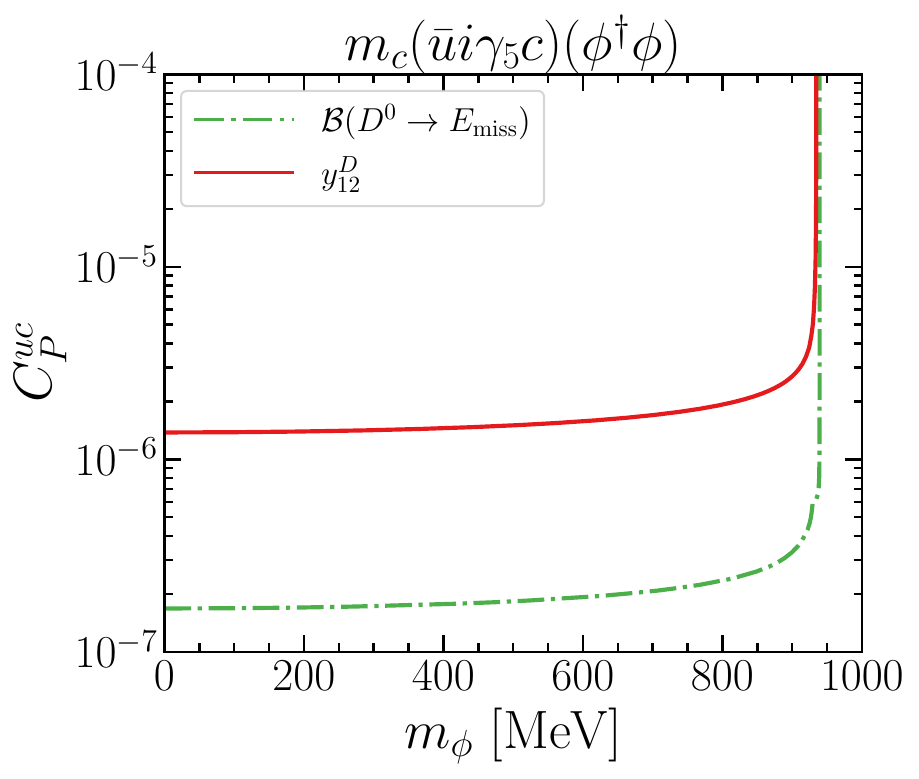}\\
     \includegraphics[width=4.7cm]{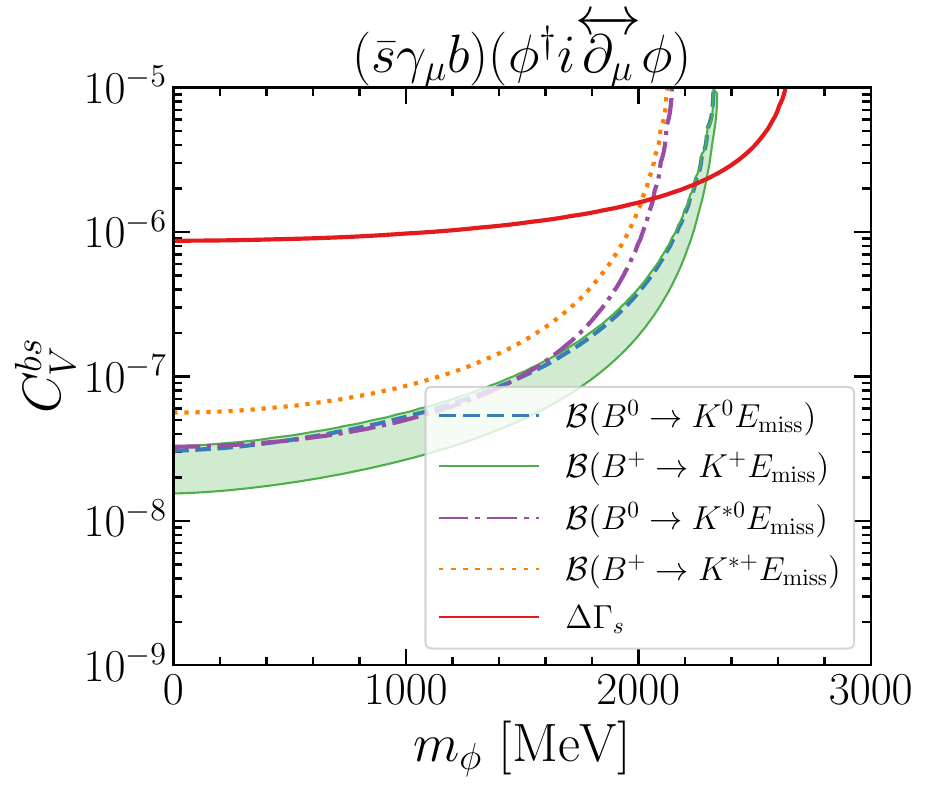}\includegraphics[width=4.7cm]{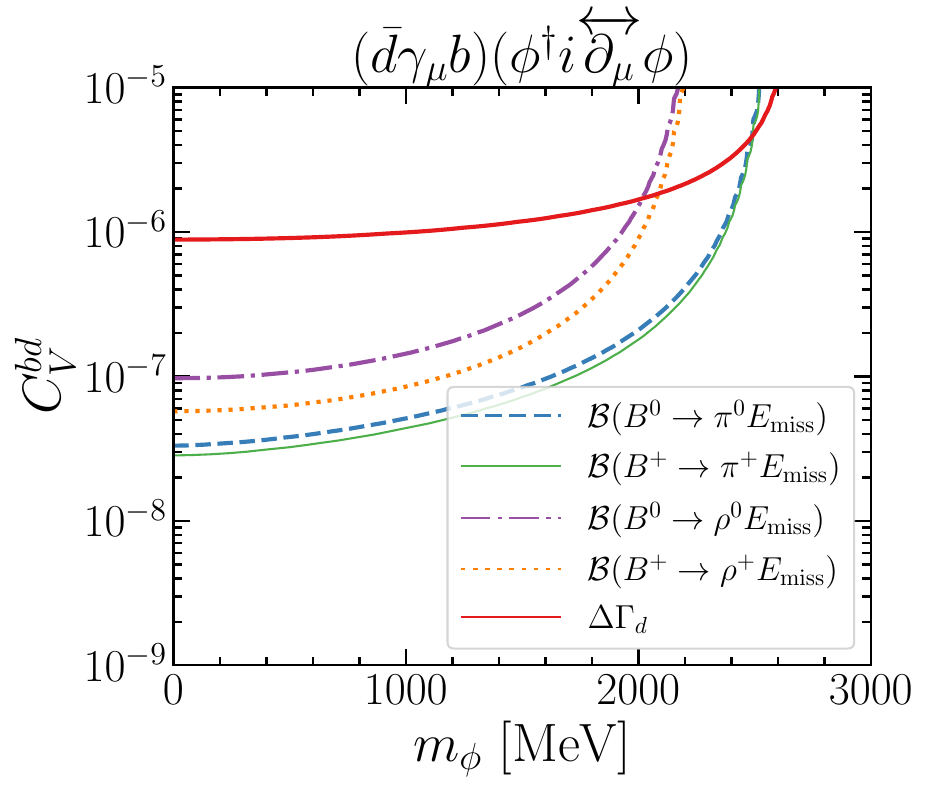}\includegraphics[width=4.7cm]{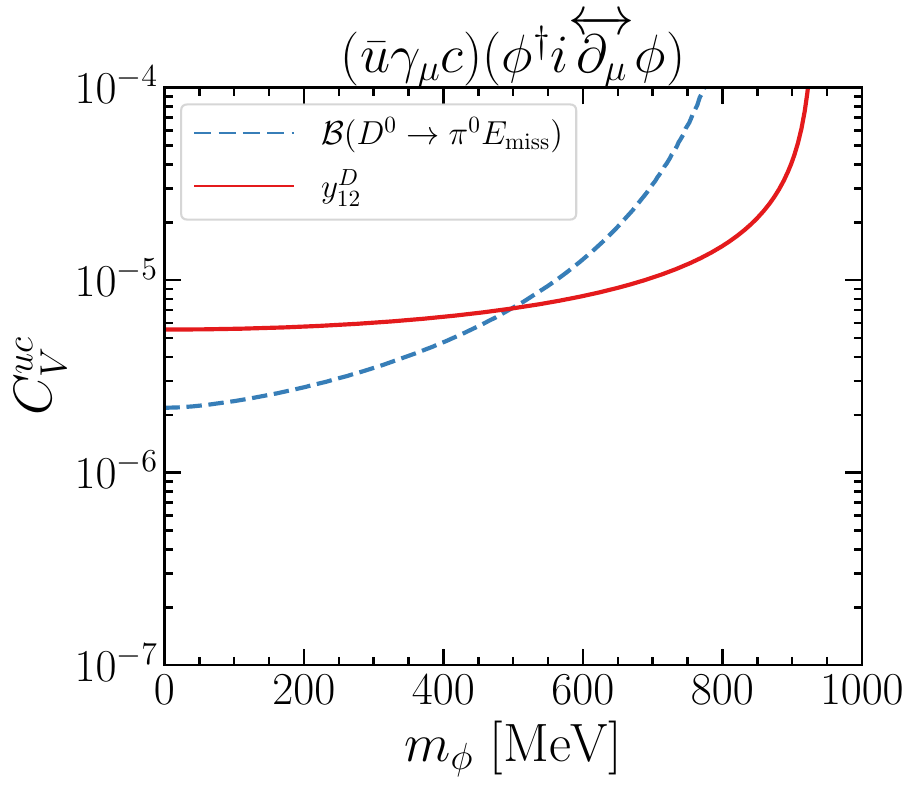}\\
     \includegraphics[width=4.7cm]{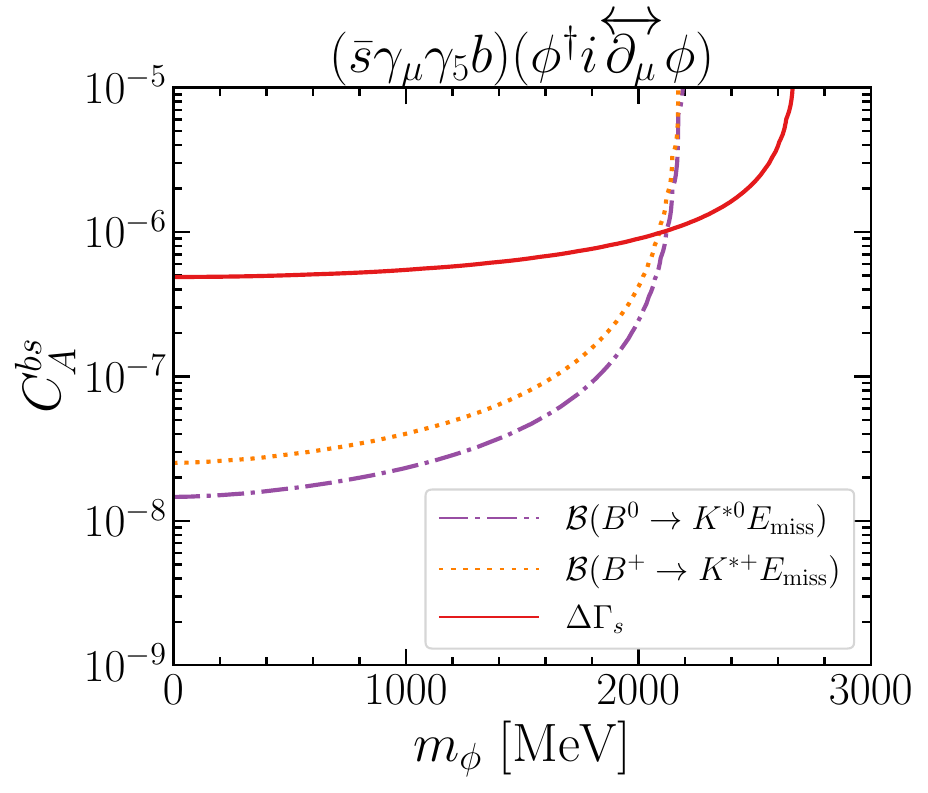}\includegraphics[width=4.7cm]{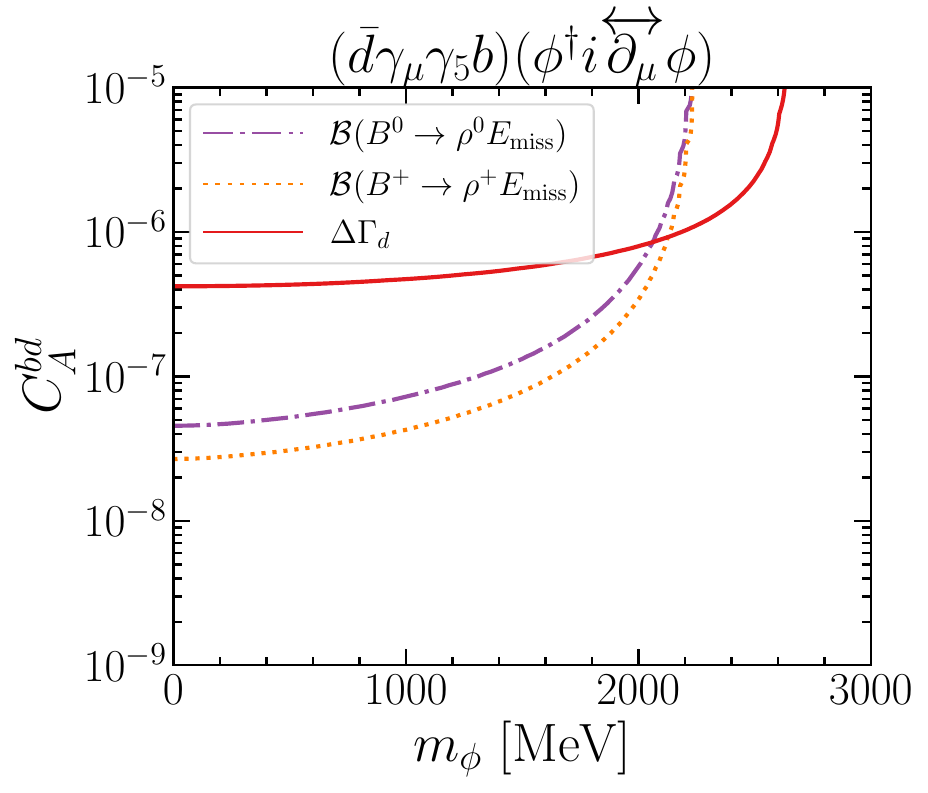}\includegraphics[width=4.7cm]{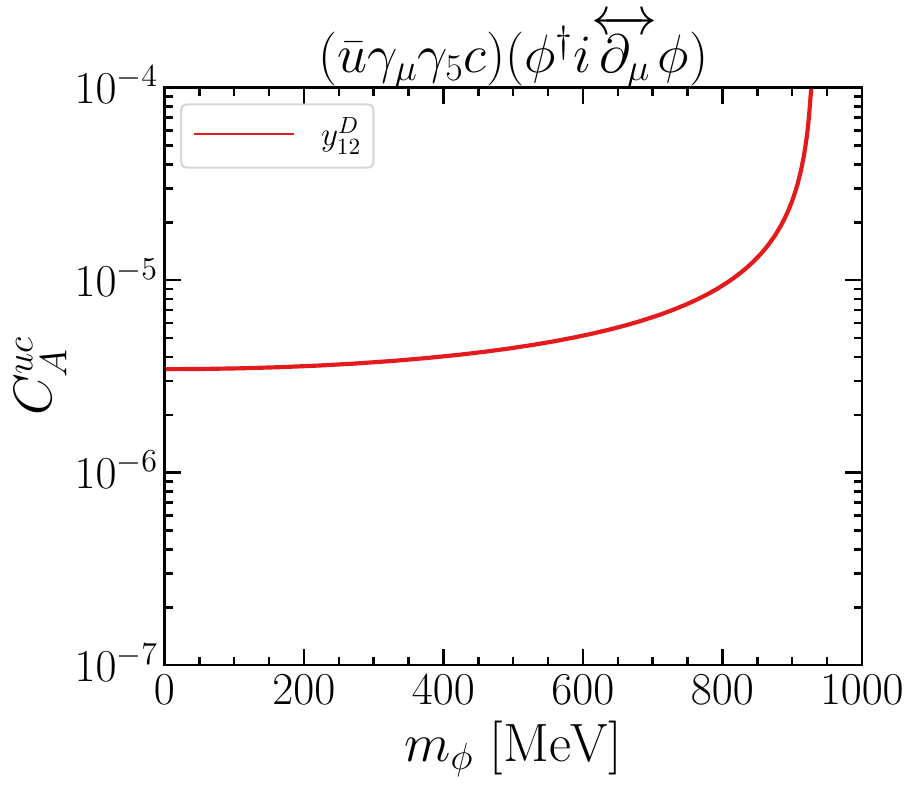}
\caption{Constraints on scalar DM operators in Eq.~\eqref{eq: scalar-ops}. See  text for details.}
\label{fig: scalar-DM}
\end{figure}

In Fig.~\ref{fig: scalar-DM}, we show constraints in the Wilson coefficient
$(C_i^{q_iQ_j})$ vs. DM mass ($m_\phi$) plane for all effective operators 
in the scalar DM scenario. We set $\Lambda = 1$ TeV, as the indirect measurements 
only constraint the combination $C_i^{q_iQ_j}/\Lambda^2$.

Note that unless mentioned otherwise explicitly all constraints
in each of plots throughout the paper correspond to exclusion curves at $90\% $ CL,
with region above the curves being ruled out.

The first column of  Fig.~\ref{fig: scalar-DM} shows constraints on
$\{\bar s b\}$ quark flavor operators   from $\Delta \Gamma_s$
(solid red line),  $\mathcal{B}(B^0\to K^0 \Emiss)$
(dashed blue line),  $\mathcal{B}(B^0\to K^{\ast 0} \Emiss)$
(dash-dotted purple line), $\mathcal{B}(B^+\to K^{\ast +} \Emiss)$
(dotted orange line), and $\mathcal{B}(B_s^0\to \Emiss)$
(dashed brown line). In contrast, the green shaded band corresponds to $90\% $ CL 
region that can explain the recent Belle II measured value of
$\mathcal{B}(B^+\to K^{ +} \Emiss)$.
Concerning branching ratio constraints, note that operators with scalar quark current
$(\bar s b)$ are insensitive to decays involving vector meson final state since the
corresponding matrix element vanishes, \emph{e.g.}, 
$\langle K^\ast| \bar s b| B \rangle =0$. Similarly, operators
with axial ($\bar s \g_\mu\g_5 b$) or pseudoscalar ($\bar s \g_5 b$) quark currents
are not sensitive to decays involving pseudoscalar meson final state  as 
$\langle K| \bar s \,\Gamma_i b| B \rangle =0$ for $\Gamma_i = \g_5$ and
$\g_\mu\g_5$.  
Overall, we find that constraints from $\Delta \Gamma_s$ are always weaker,
by an order or more, compared to those from decay modes.
However, note that the region beyond DM mass
$m_\phi \ge (m_B - m_{K^{(\ast)}})/2$ becomes kinematically inaccessible in three-body decays.
This region, except for the operator $\mathcal{O}_{P}^{sb}$, is only constrained from $\Delta \Gamma_s$.

In the second column of Fig.~\ref{fig: scalar-DM} we show constraints on
$\{\bar d b\}$ quark flavors operators from $\Delta \Gamma_d$
(solid red line),  $\mathcal{B}(B^0\to \pi^0 \Emiss)$
(dashed blue line), $\mathcal{B}(B^+\to \pi^+ \Emiss)$
(solid green line), $\mathcal{B}(B^0\to \rho^{0} \Emiss)$
(dash-dotted purple line), $\mathcal{B}(B^+\to \rho^{+} \Emiss)$
(dotted orange line), and $\mathcal{B}(B^0\to \Emiss)$
(dashed brown line). The pattern of constraints is similar
to those discussed above. But now since allowed phase spaces of $B^0\to \Emiss$ and
$B\to \pi \Emiss$ decays are relatively larger, their constraints cover the 
whole region of $\Delta \Gamma_d$ bound.

In the third column of Fig.~\ref{fig: scalar-DM} we show constraints on operators
with $\{\bar u c\}$ quark flavors. The bound from $y_{12}^D$ (\emph{i.e.,} $y_{12}^D \le 0.641\%$)
is shown as solid red line, whereas bounds from branching ratios of
$D^0\to \pi^0 \Emiss$ and  $D^0\to \Emiss$ are shown as dashed blue line and
dash-dotted green line, respectively. The relevance of including 
lifetime difference constraint is particularly visible now.
Because only a couple of charm decay modes have been measured so far,
not all effective operators can be constrained from them.
The lifetime difference constraint is the sole constraint on such operators,
\emph{e.g.}, $(\bar u \g_\mu\g_5 c)(\phi^\dagger i\overleftrightarrow{\partial_\mu}\phi)$ (see the last plot in Fig.~\ref{fig: scalar-DM}).
Furthermore, we also note that $y_{12}^D$ constraint can in general also compete
with branching ratio constraints. For example, in the plot showing constraints
on the Wilson coefficient of operator 
$(\bar u \g_\mu c)(\phi^\dagger i\overleftrightarrow{\partial_\mu}\phi)$,
the $y_{12}^D$ constraint for $m_\phi\gtrsim 500$ MeV is stronger
than $\mathcal{B}(D^0\to \pi^0\Emiss)$ constraint.

\begin{figure}[t!]
\centering
     \includegraphics[width=4.6cm]{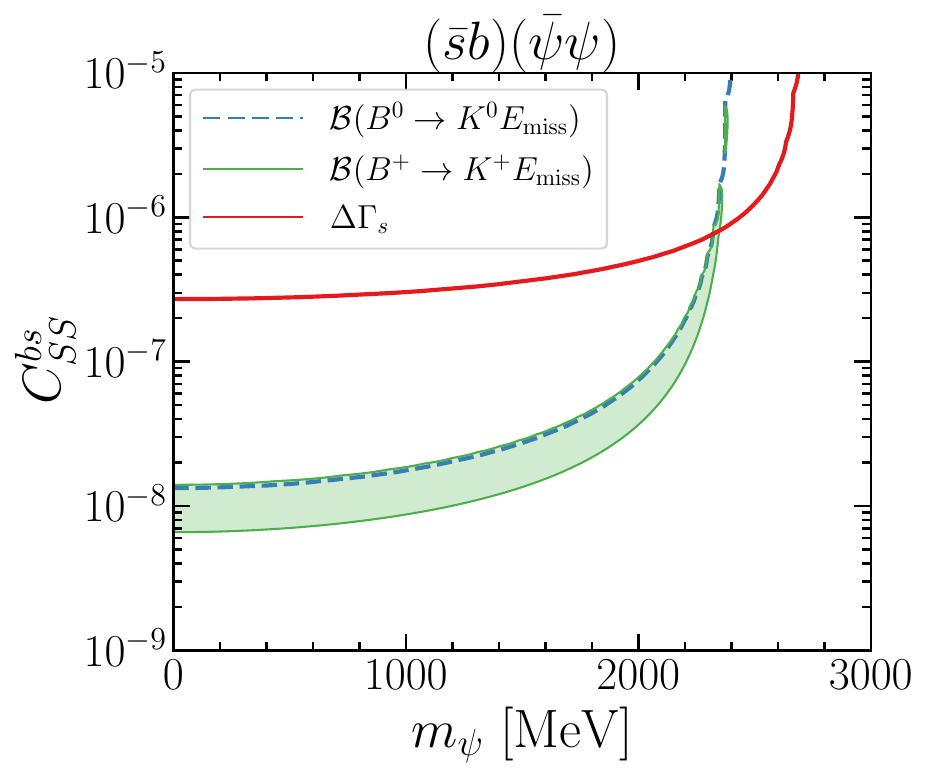}\includegraphics[width=4.6cm]{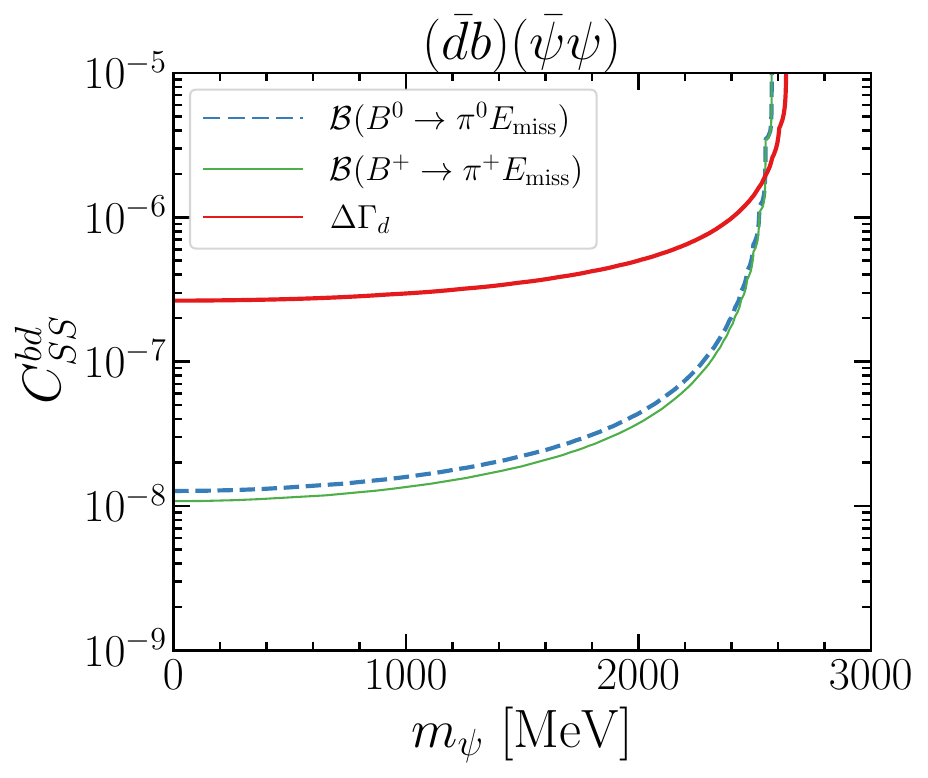}\includegraphics[width=4.6cm]{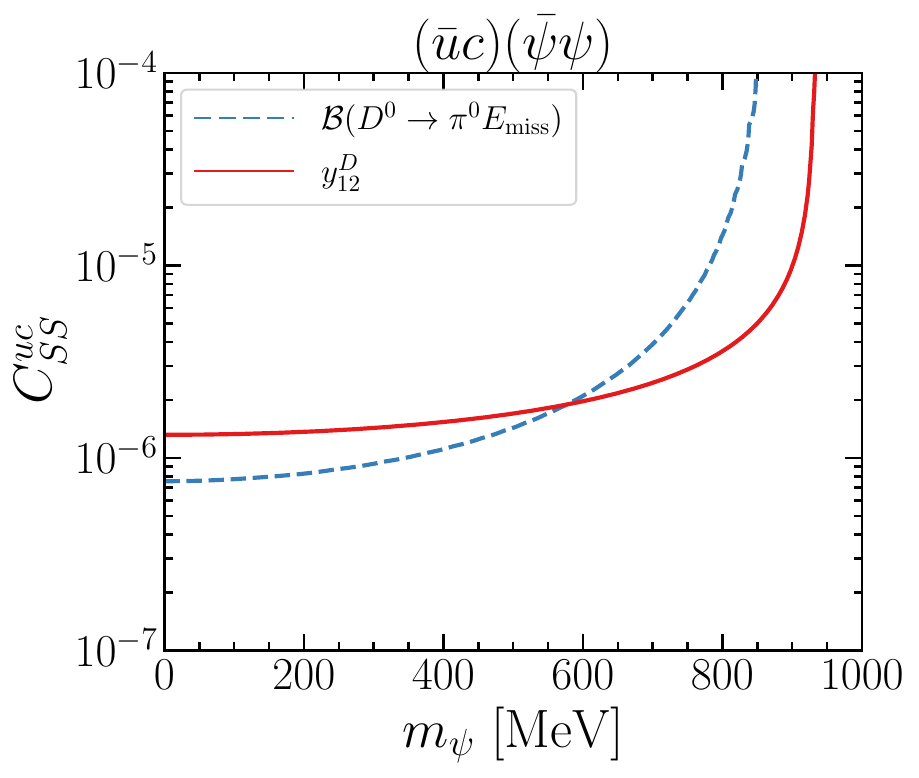}
     \includegraphics[width=4.6cm]{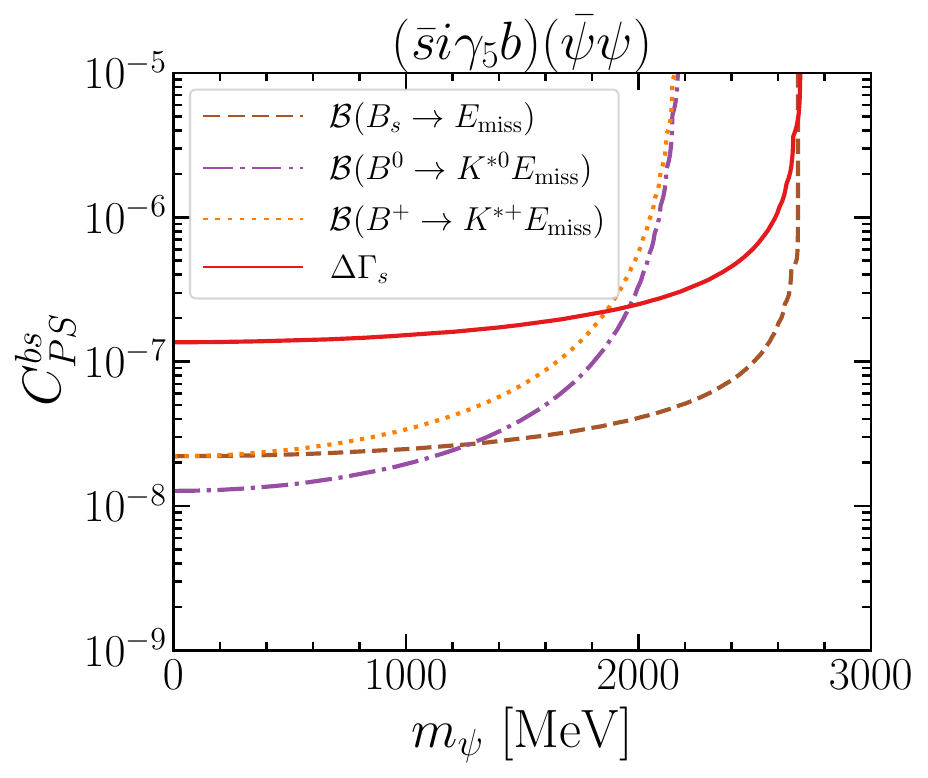}\includegraphics[width=4.6cm]{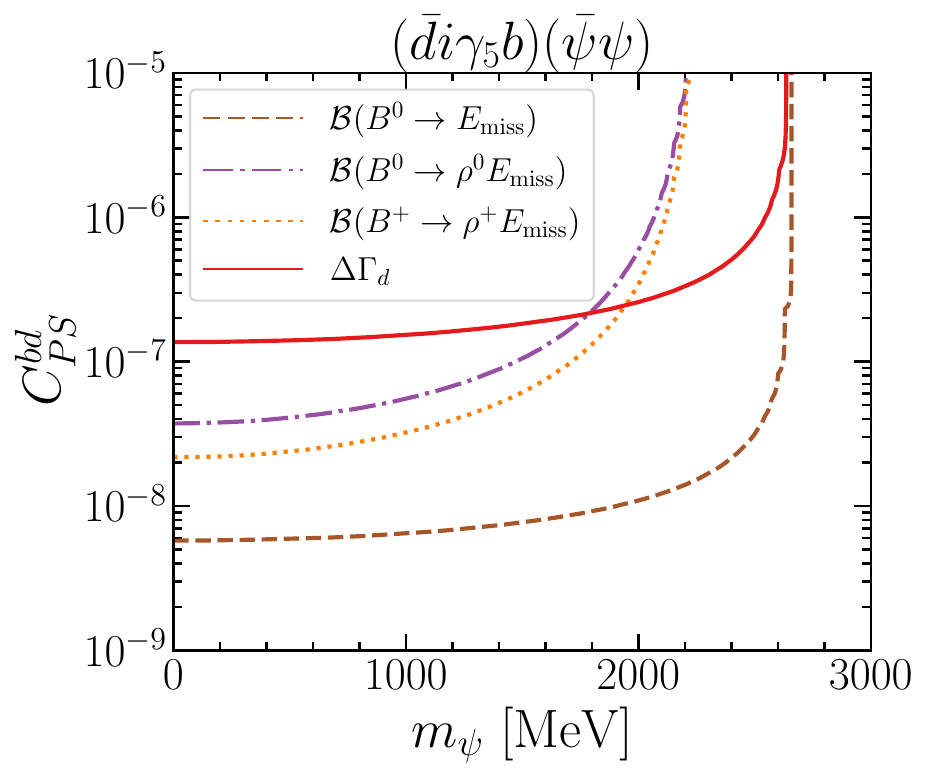}\includegraphics[width=4.6cm]{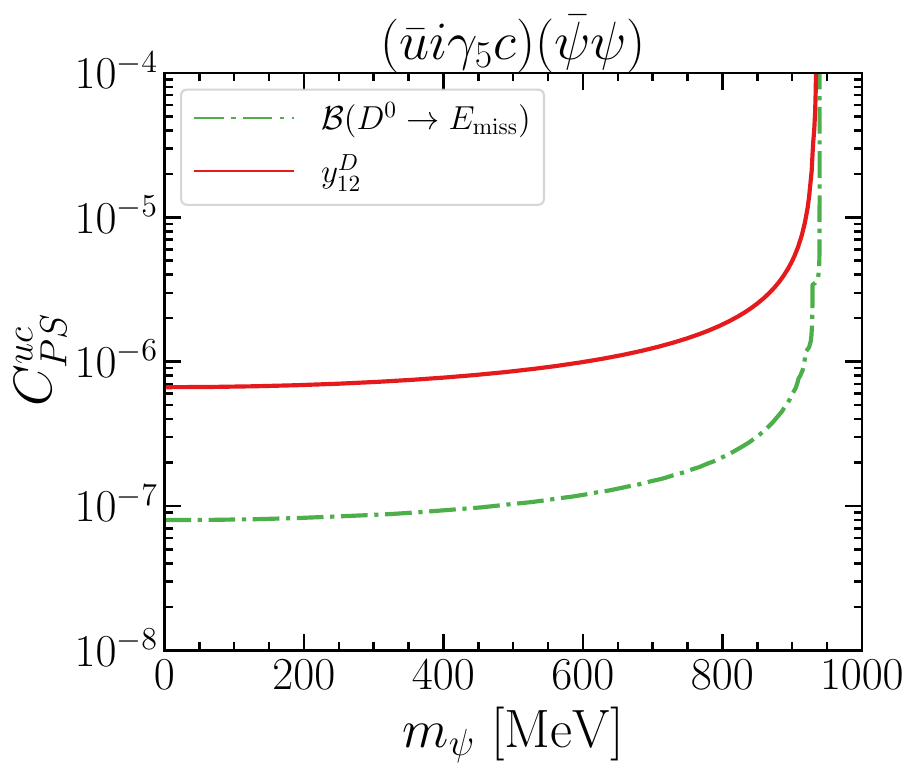}
    \includegraphics[width=4.6cm]{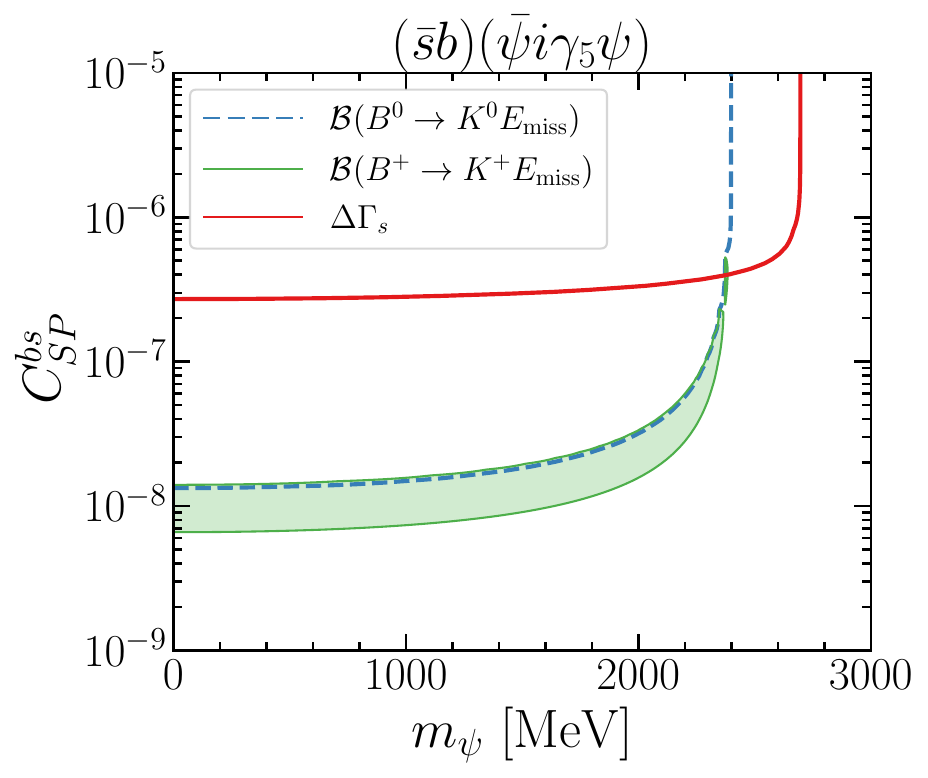}\includegraphics[width=4.6cm]{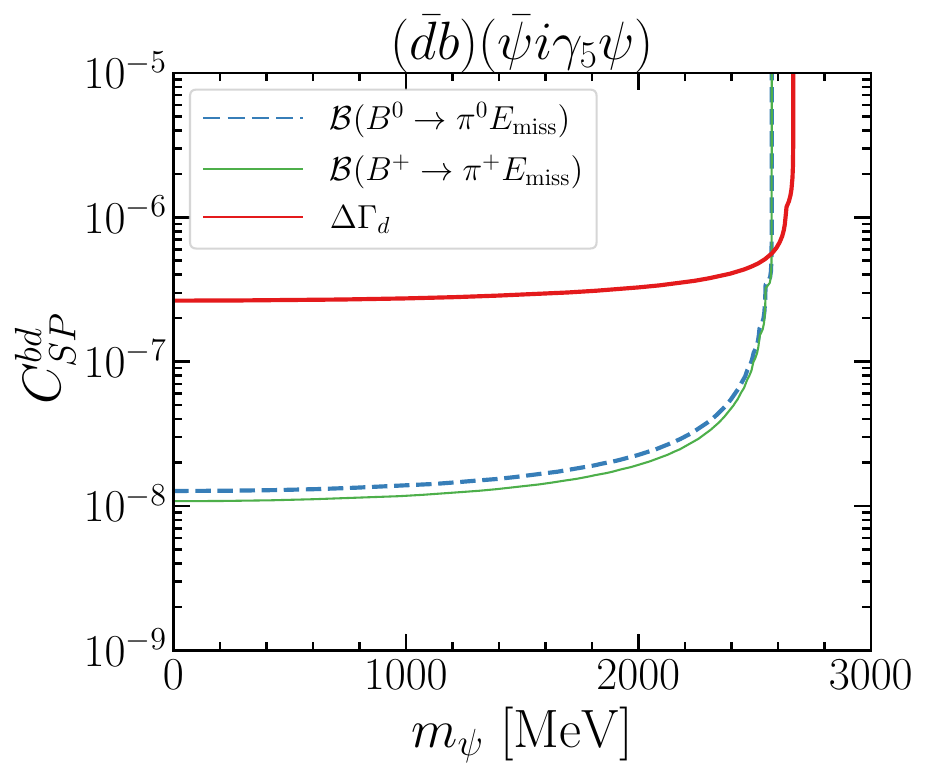}\includegraphics[width=4.6cm]{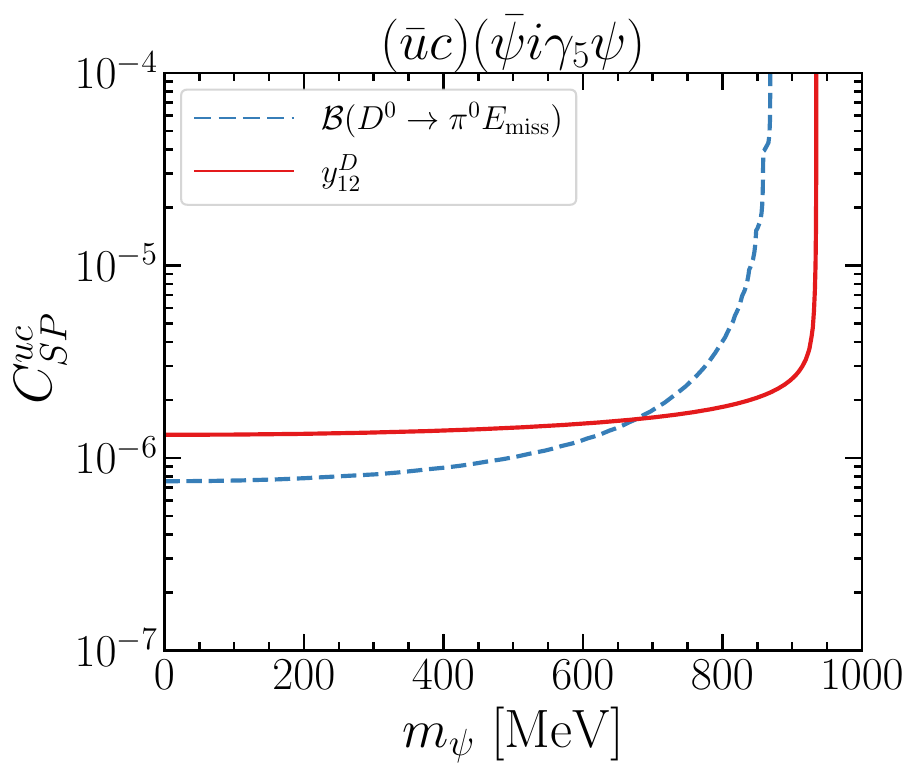}
     \includegraphics[width=4.6cm]{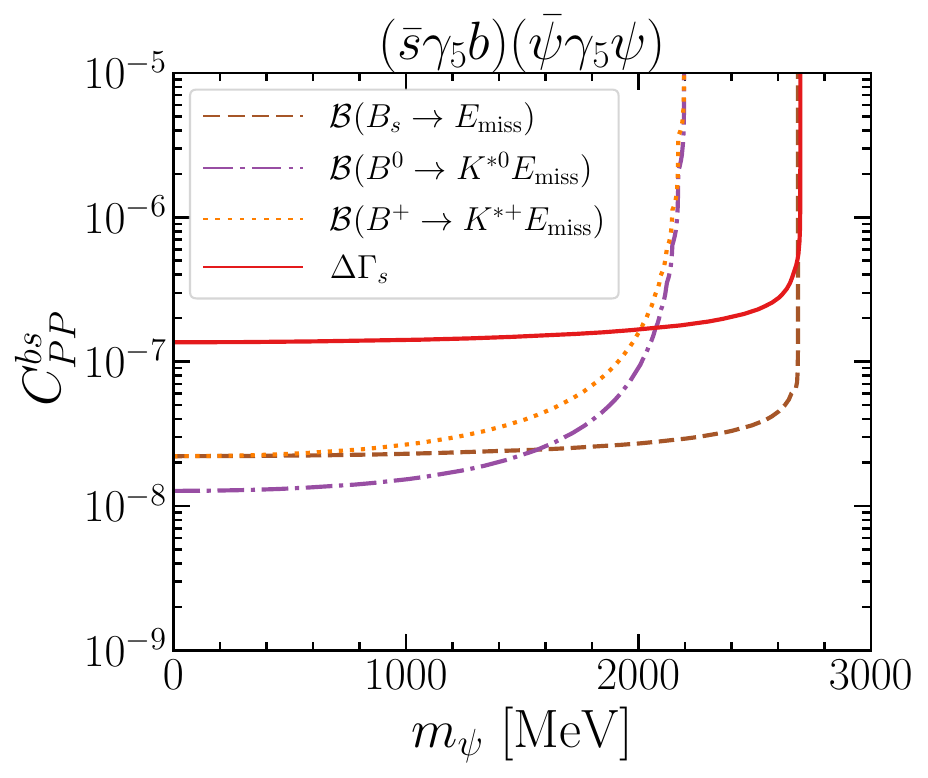}\includegraphics[width=4.6cm]{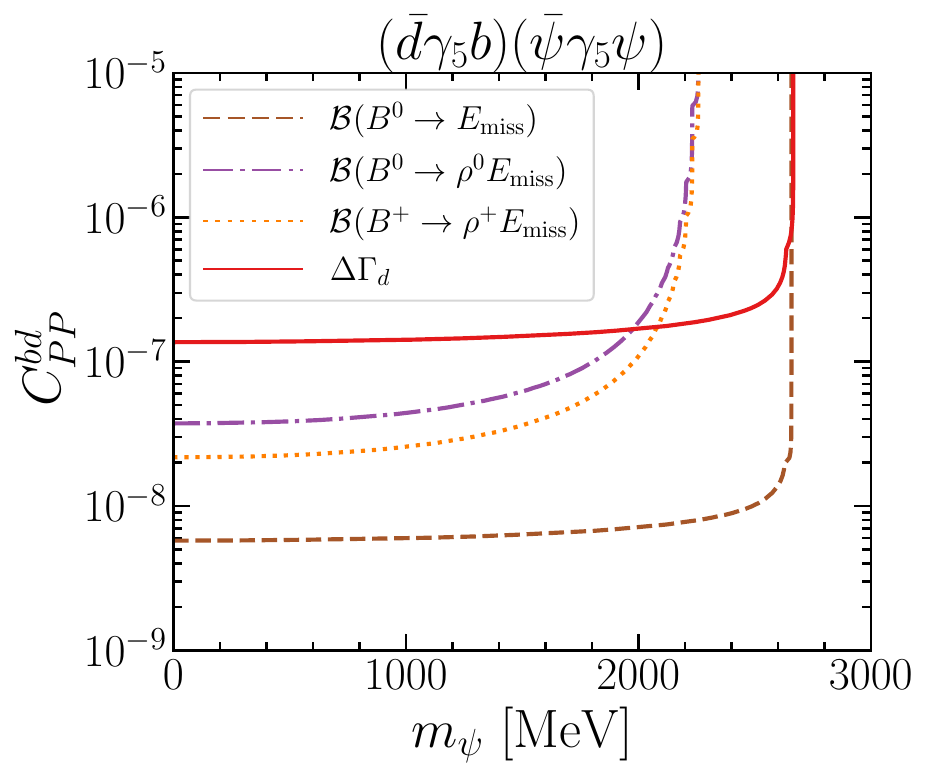}\includegraphics[width=4.6cm]{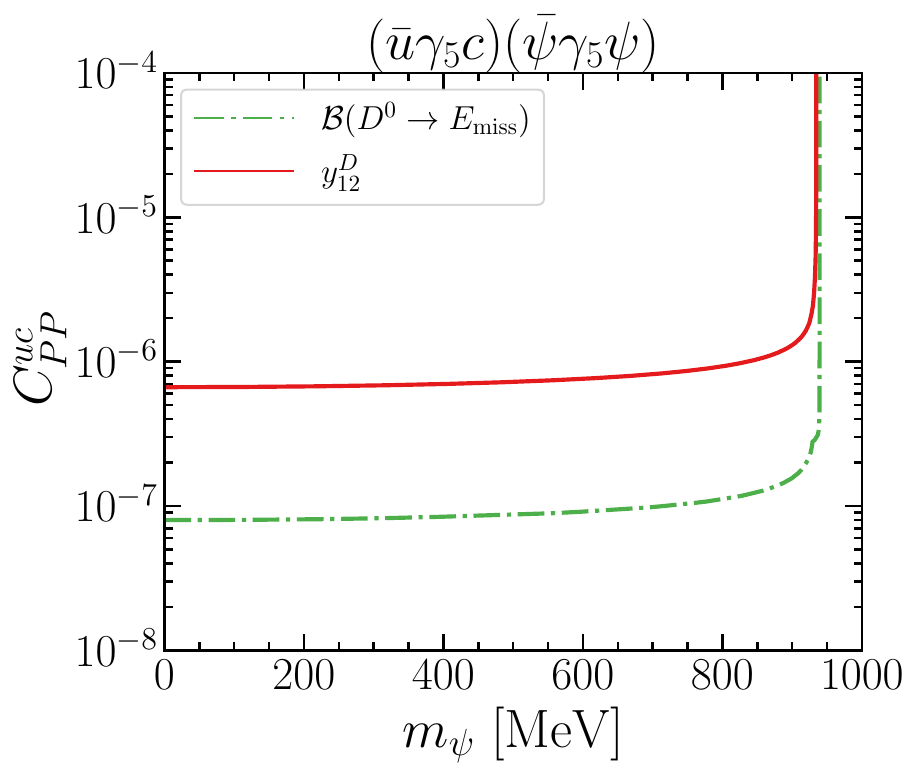}
\caption{Constraints on fermionic DM operators with (pseudo) scalar currents.}
\label{fig: fermion-DM-scalar}
\end{figure}
\subsection{Constraints on fermionic DM}\label{subsec: results-fermion}
\begin{figure}[t!]
\centering
     \includegraphics[width=4.6cm]{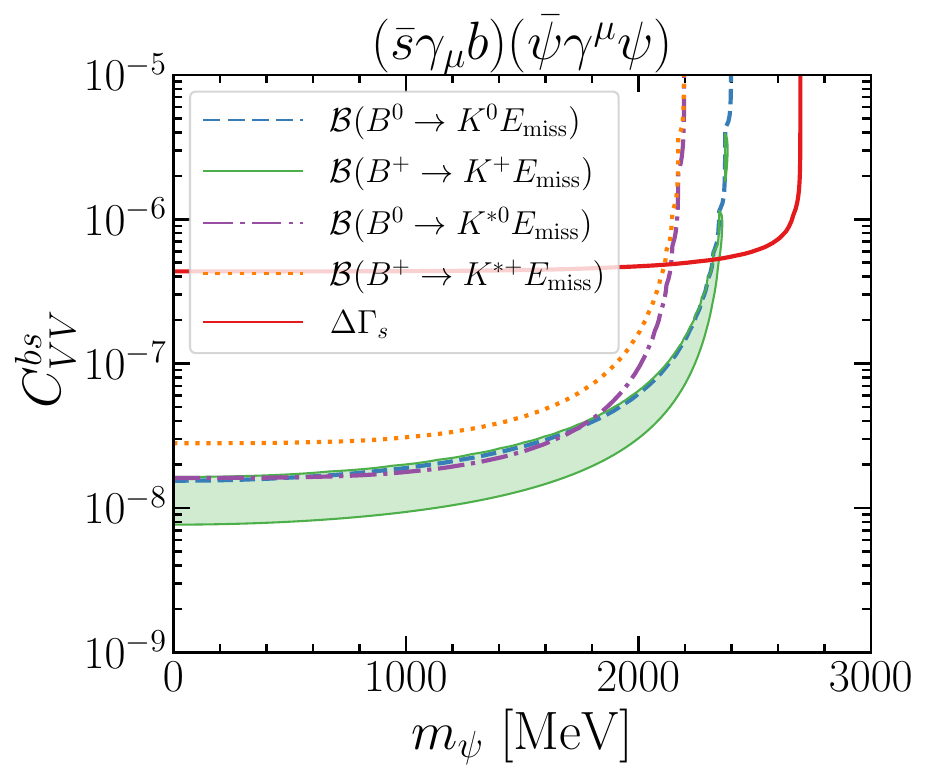}\includegraphics[width=4.6cm]{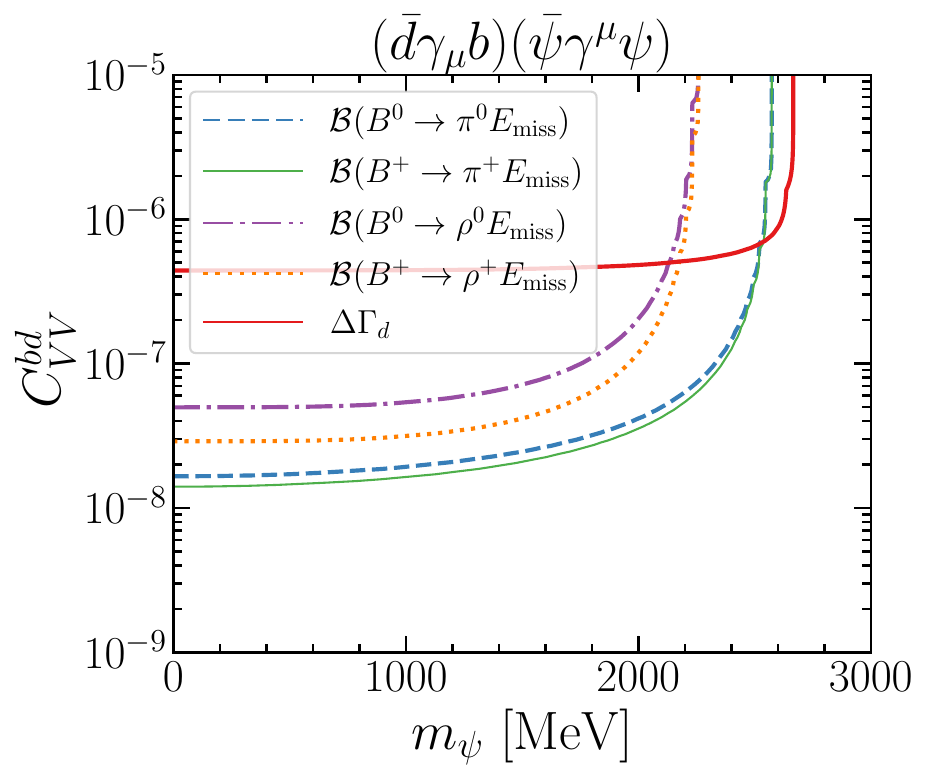}\includegraphics[width=4.6cm]{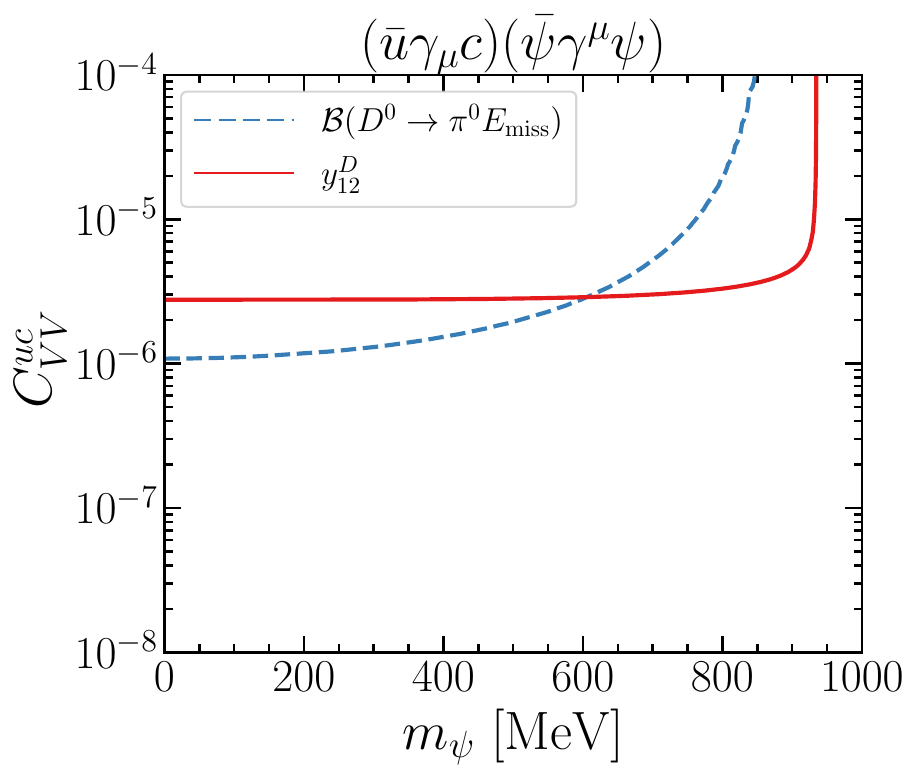}
     \includegraphics[width=4.6cm]{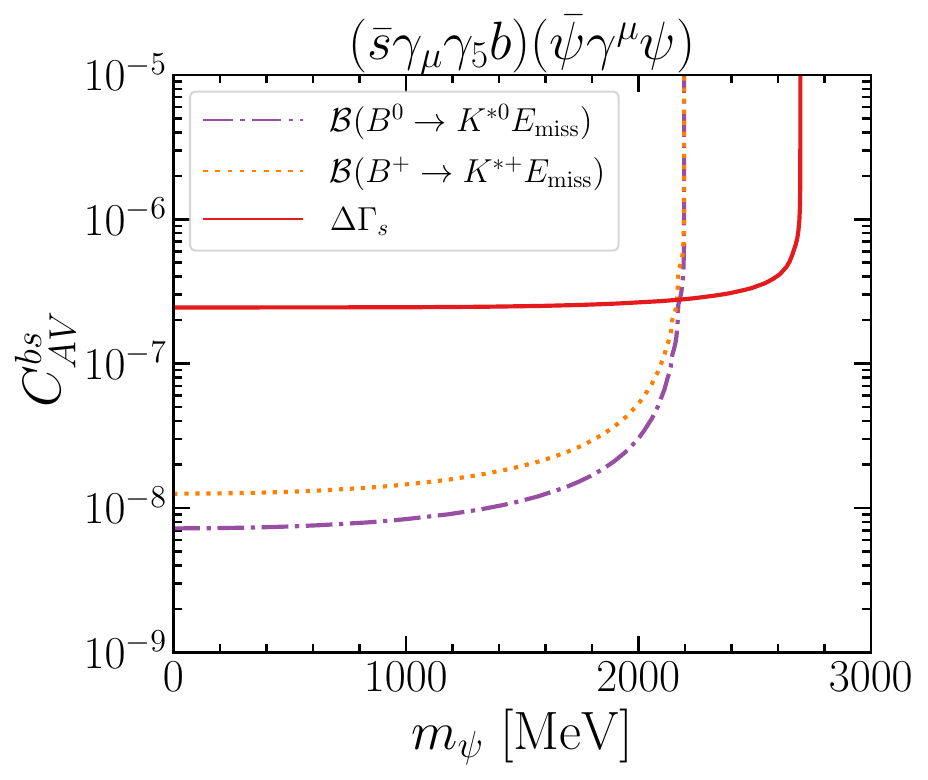}\includegraphics[width=4.6cm]{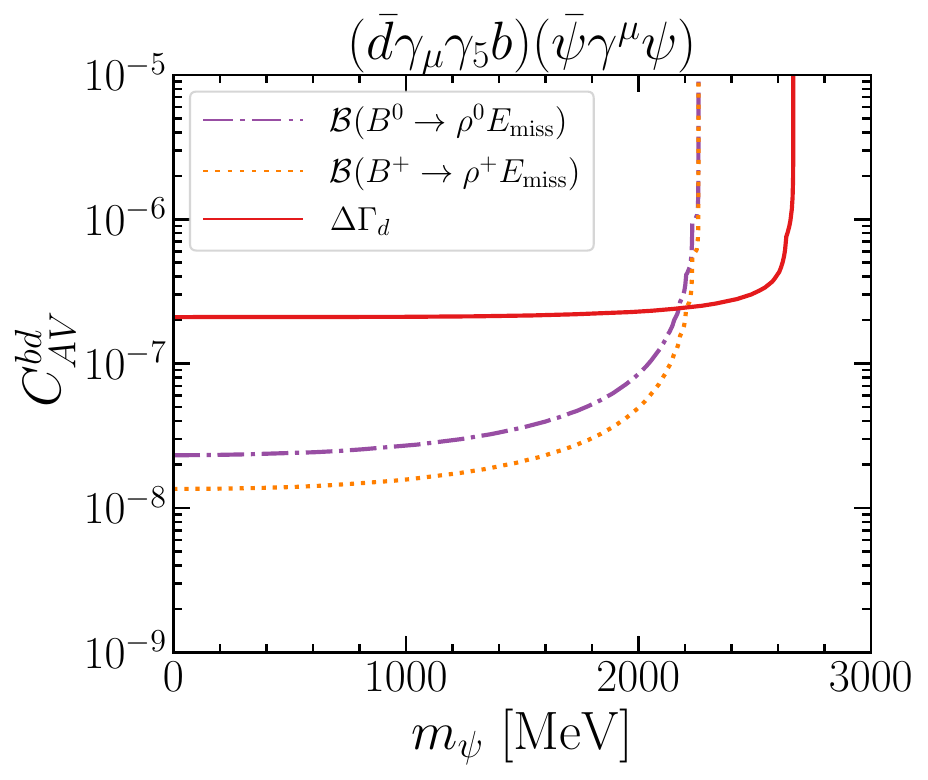}\includegraphics[width=4.6cm]{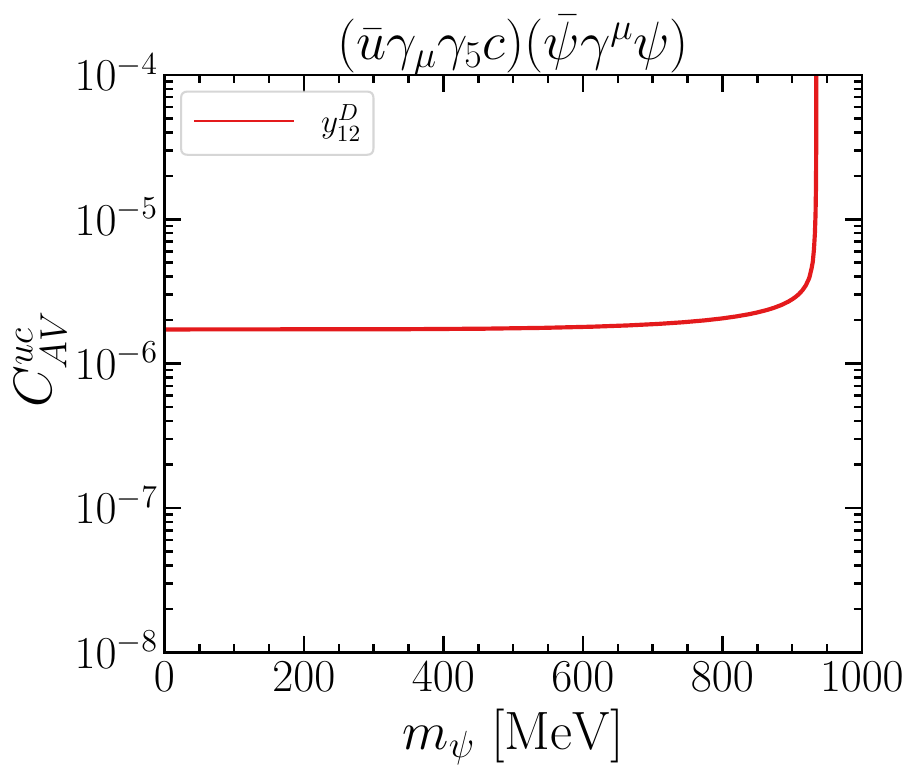}
     \includegraphics[width=4.6cm]{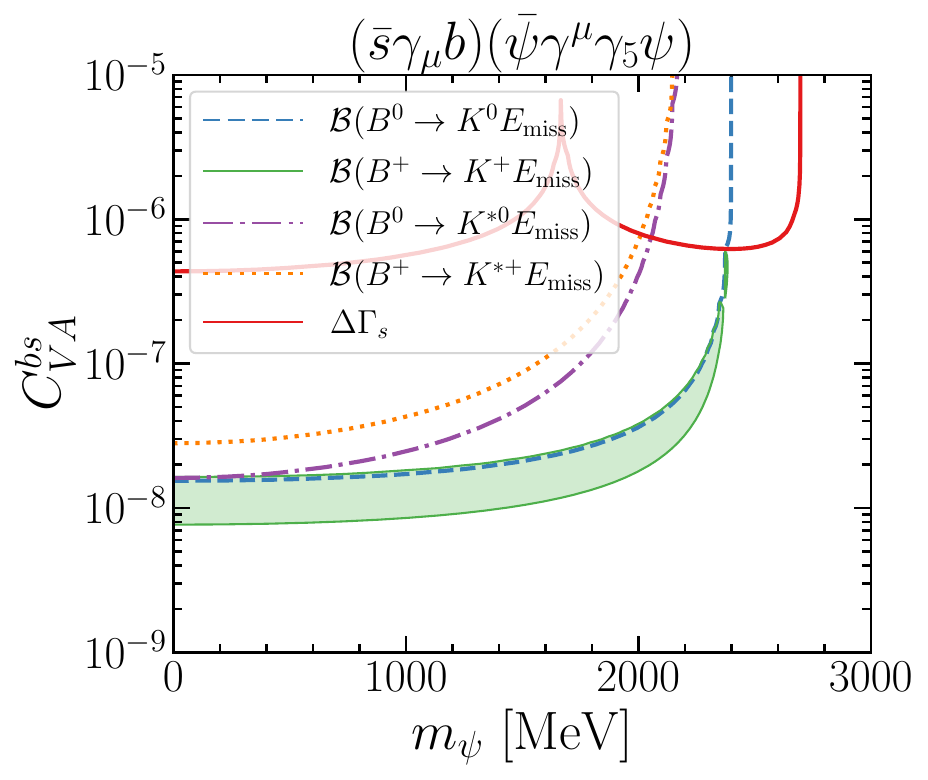}\includegraphics[width=4.6cm]{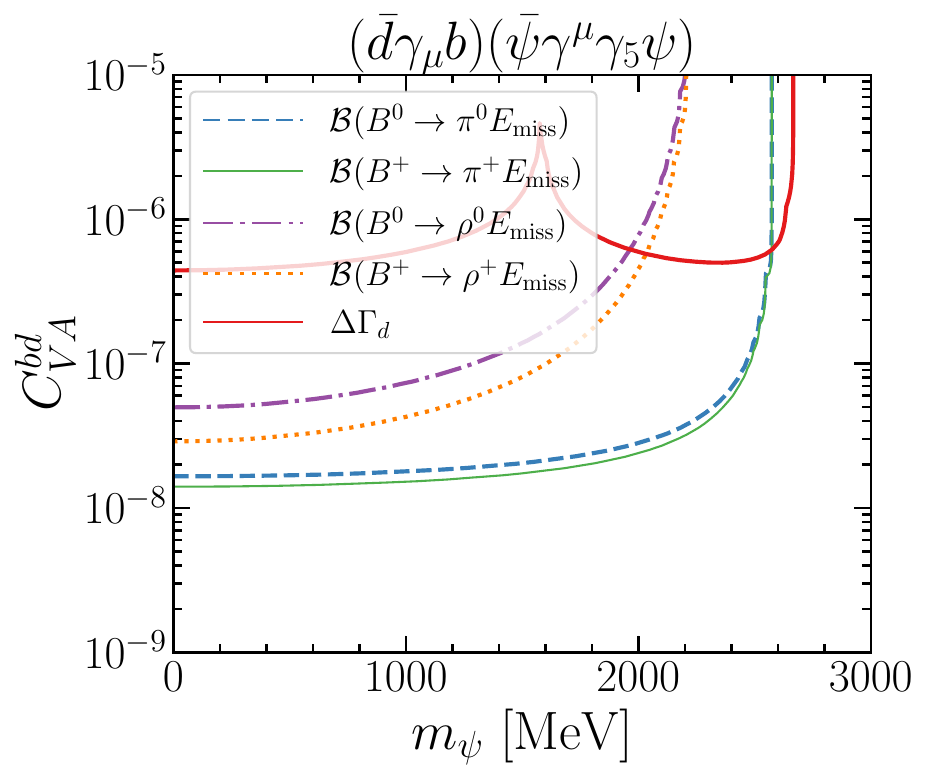}\includegraphics[width=4.6cm]{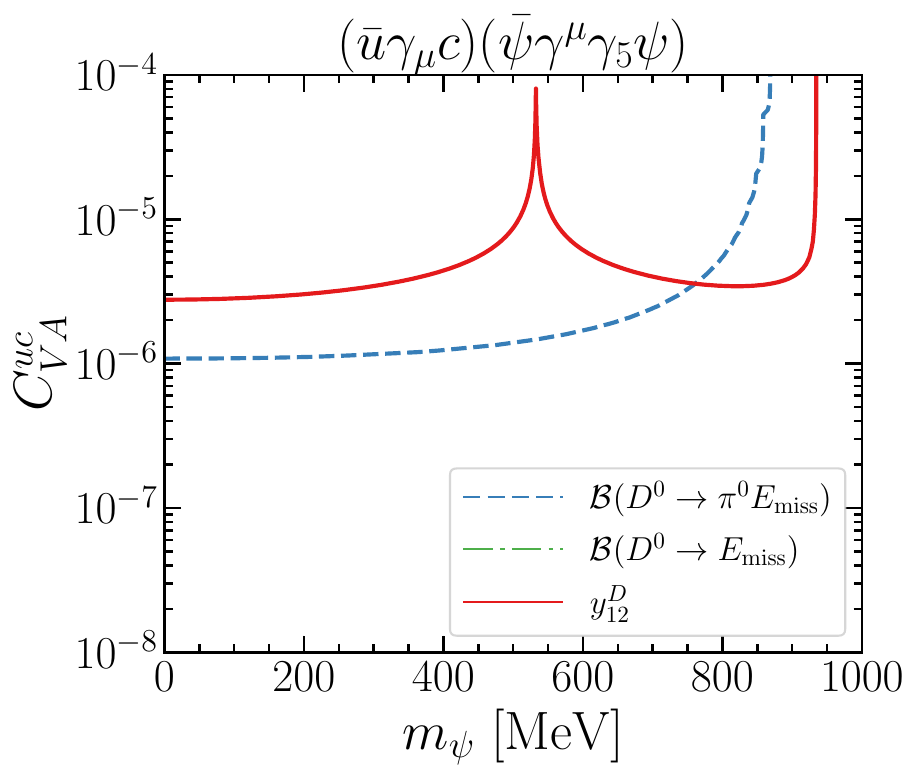}
     \includegraphics[width=4.6cm]{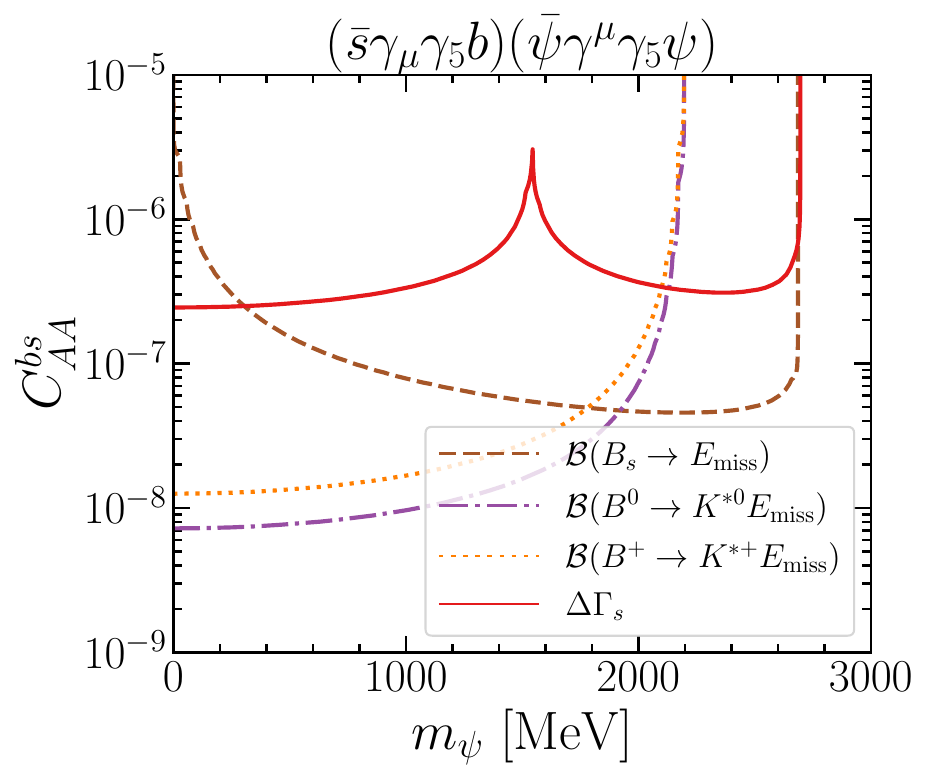}\includegraphics[width=4.6cm]{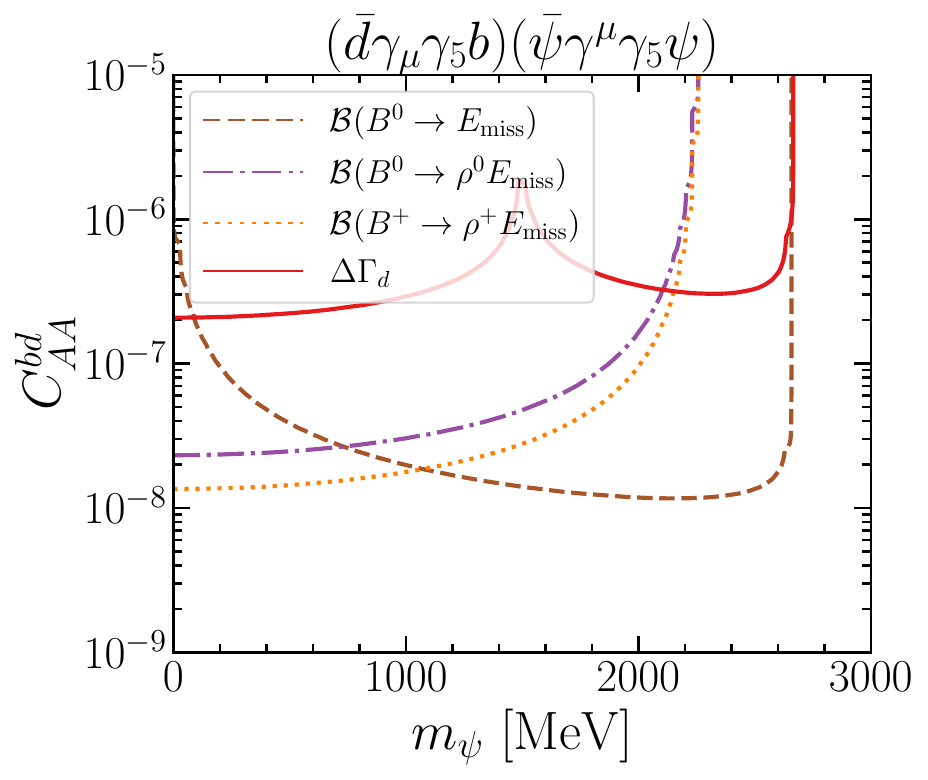}\includegraphics[width=4.6cm]{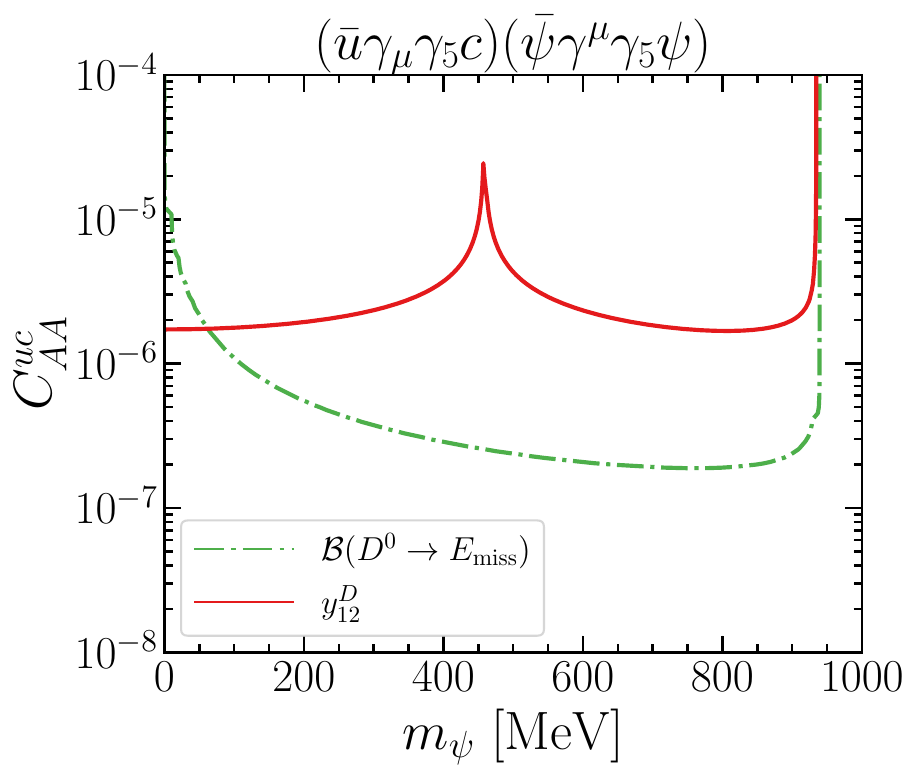}
\caption{Constraints on fermionic DM operators with (axial) vector currents.}
\label{fig: fermion-DM-vector}
\end{figure}

\begin{figure}[t!]
\centering
     \includegraphics[width=4.6cm]{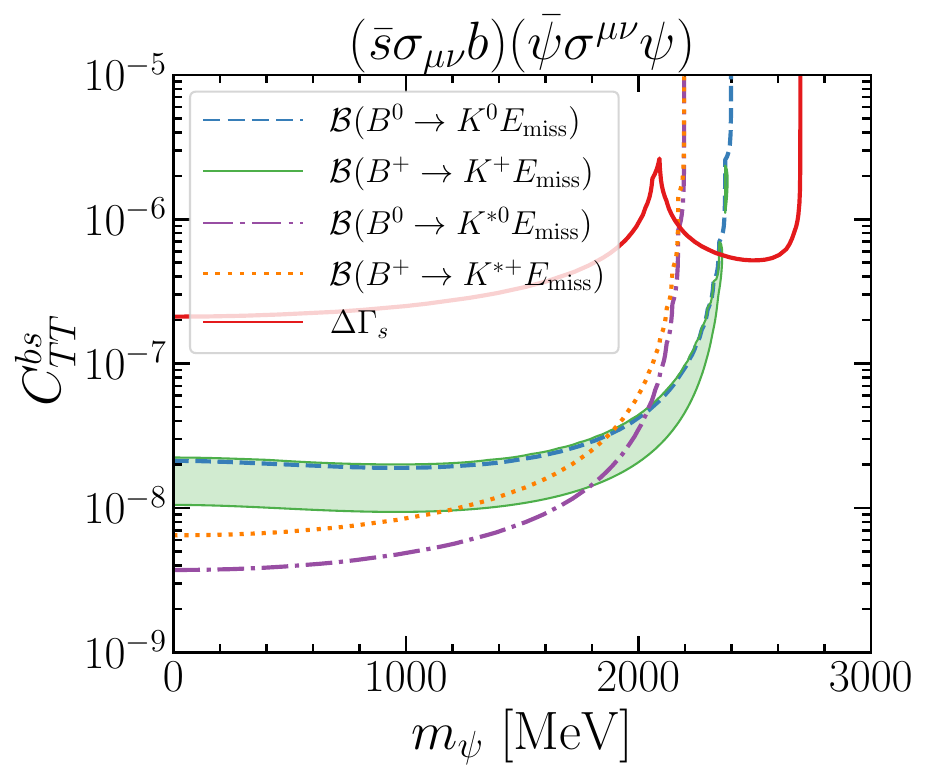}\includegraphics[width=4.6cm]{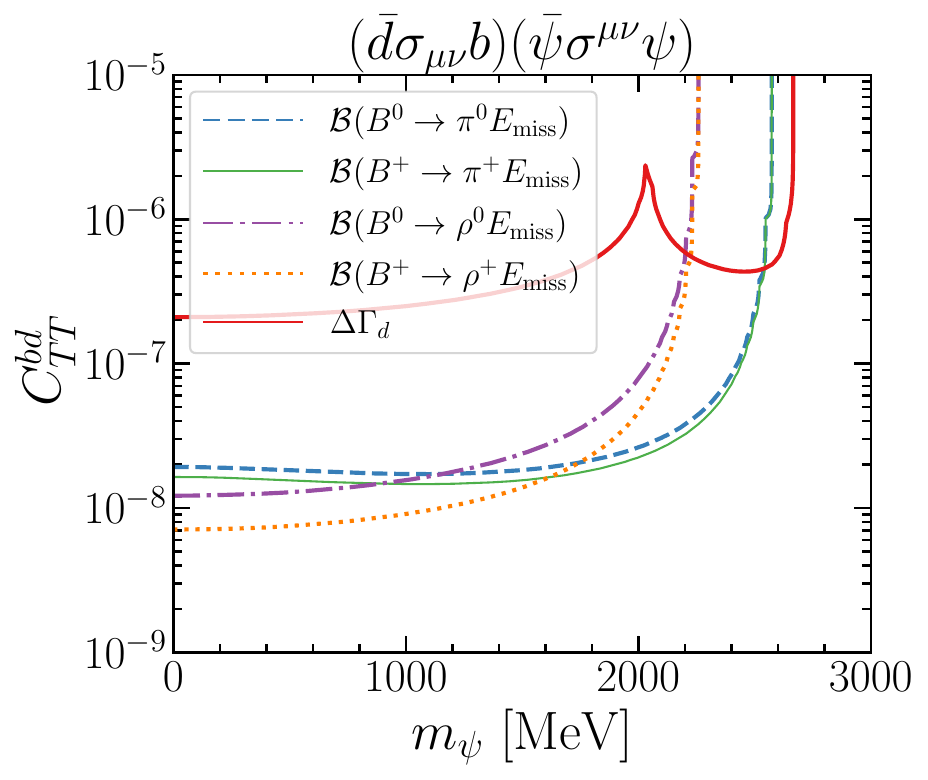}\includegraphics[width=4.6cm]{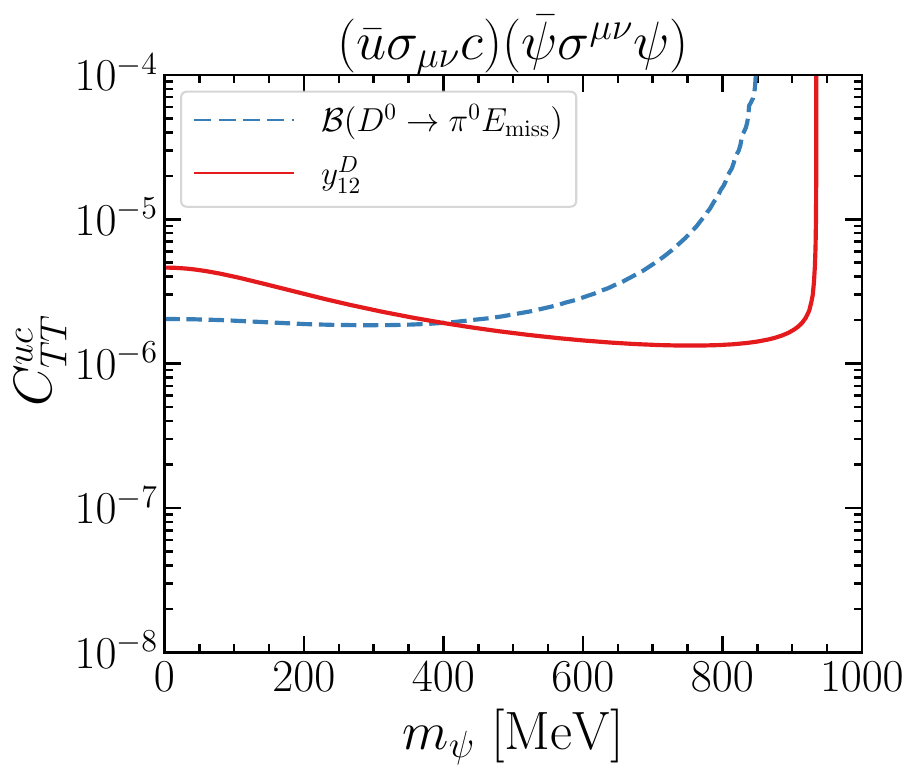}
     \includegraphics[width=4.6cm]{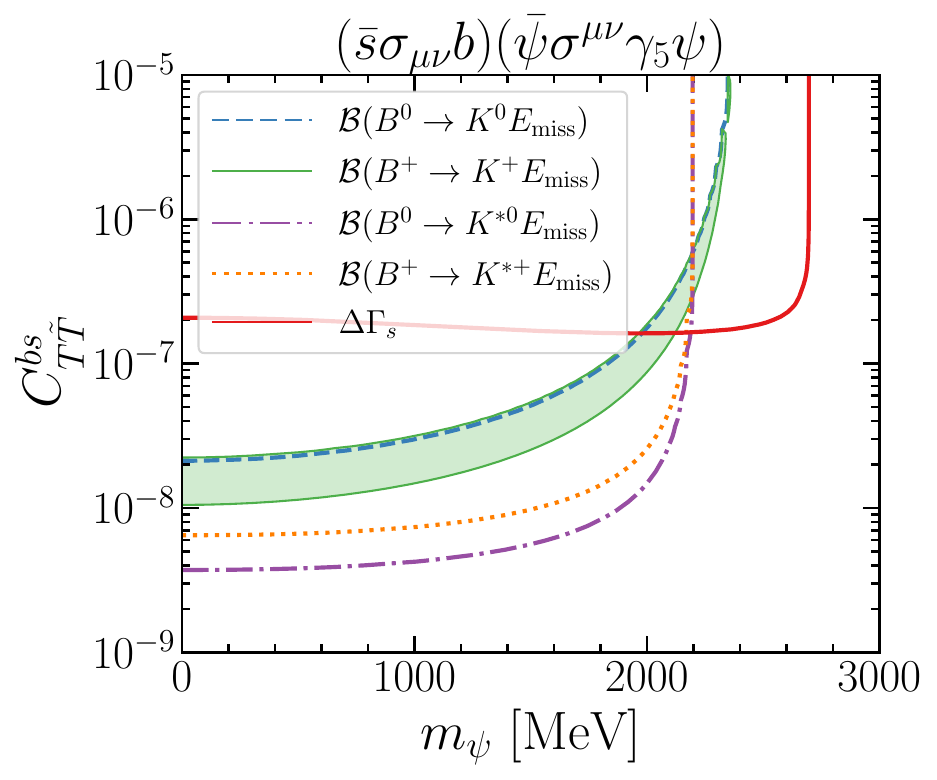}\includegraphics[width=4.6cm]{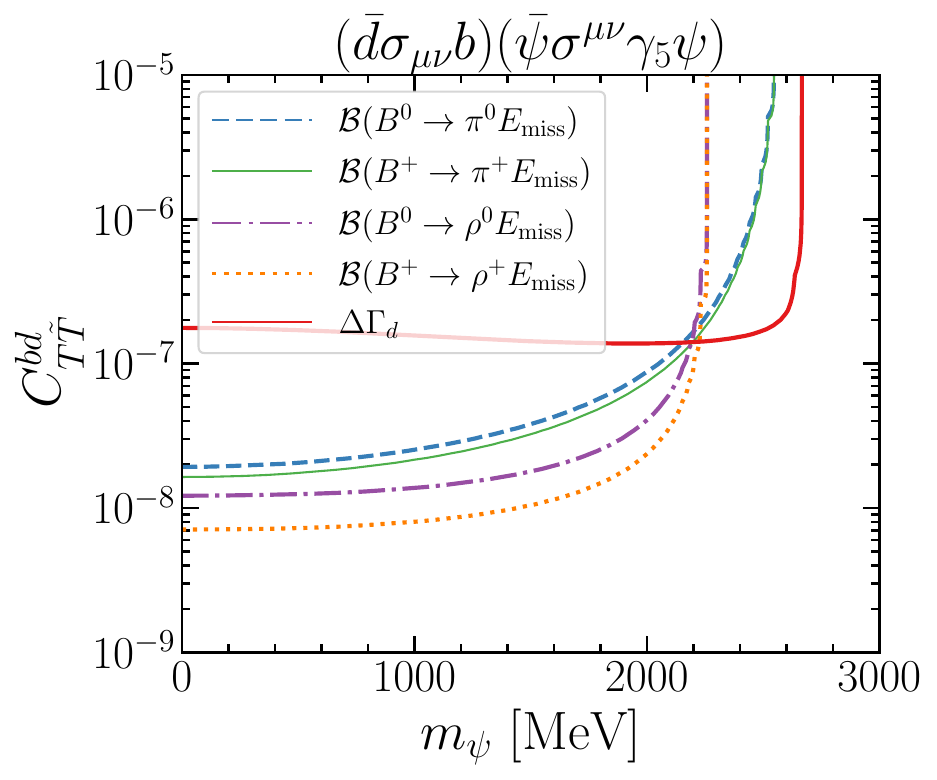}\includegraphics[width=4.6cm]{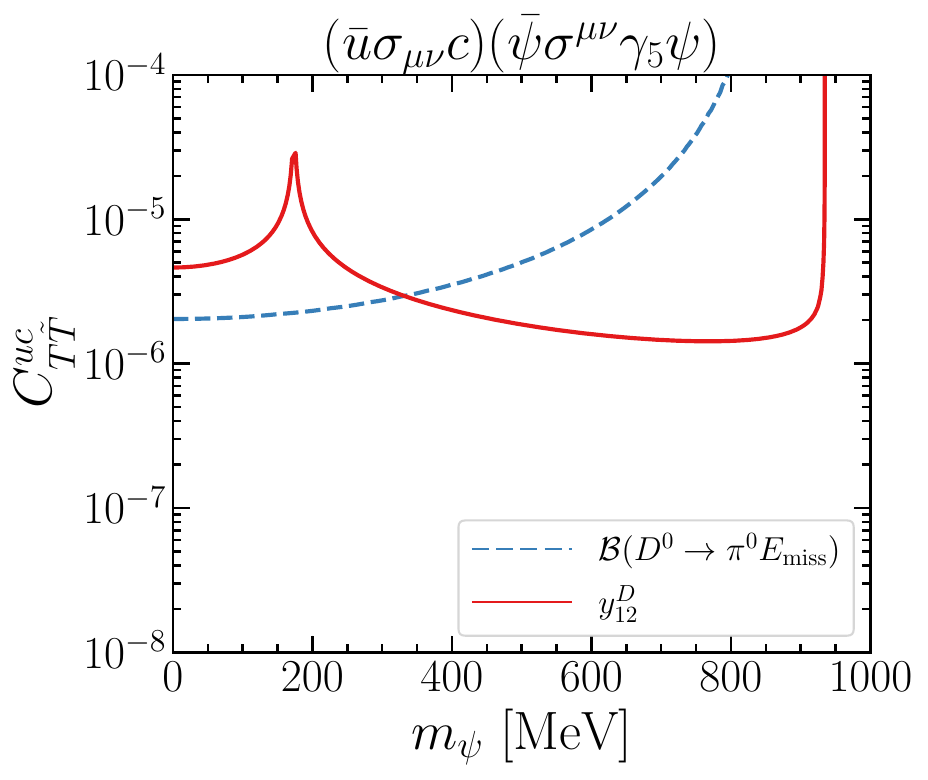}
\caption{Constraints on fermionic DM operators with tensor currents.}
\label{fig: fermion-DM-tensor}
\end{figure}

In Fig.~\ref{fig: fermion-DM-scalar}, \ref{fig: fermion-DM-vector}, and
\ref{fig: fermion-DM-tensor} we show combined constraints from lifetime difference
and decay modes on fermionic DM operators.
The color and style of visualization of constraints is consistent with
Fig.~\ref{fig: scalar-DM}. 

Fig.~\ref{fig: fermion-DM-scalar} displays constraints
on operators with scalar and pseudoscalar quark currents. The operators with
$b$ quark are constrained far better from various $B$ meson decays than
$\Delta\Gamma_{s, d}$ but the latter becomes important once 
DM mass is too large to be produced in decays.
For operators involving charm quark, $D^0 \to \Emiss$ and $D^0\to \pi^0\Emiss$
decays constrain operators with $\bar u \g_5 c$ and $\bar u c$ quark
currents, respectively; the bound from $D^0 \to \Emiss$ is always
stronger than $y_{12}^D$ bound for
all $m_\psi$ values, whereas bound from $D^0\to \pi^0\Emiss$ is slightly better than $y_{12}^D$ bound for small $m_\psi$, but gets weaker as $m_\psi$ increases and
is overtaken by $y_{12}^D$ for $m_\psi\gtrsim 600$ MeV.

Fig.~\ref{fig: fermion-DM-vector} shows constraints
on operators involving vector and axial quark currents. The constraints from
$B$ meson decays and $\Delta\Gamma_{s, d}$ have similar behavior
as discussed previously.
However, for charm quark related operators,
$y_{12}^D$ again becomes very important observable as can be seen from the
third column of the figure. In particular, we make two important observations.
First,  the operator $(\bar u \g_\mu \g_5c)(\bar \psi \g^\mu \psi)$
is not contained from charm decays at present;
we are therefore able to constrain this operator for the first time using
$y_{12}^D$ data.
The second point concerns constraints on operator
$(\bar u \g_\mu \g_5c)(\bar \psi \g^\mu \g_5\psi)$. 
As noted from the last plot, generally
the bound from invisible charm decay $D^0\to \Emiss$ is the leading one,
but since the corresponding decay rate is proportional to $m_\psi$,
this decay does not provide any constraint in the vicinity of $m_\psi\to 0$.
Therefore, $y_{12}^D$, which is finite, is crucial to constrain this region.
Note that lifetime difference constraint on operators involving axial DM current
$(\bar \psi \g^\mu\g_5\psi)$ (plots in the last two rows) show ``kinks'',
which stem from cancellation between various contributions to width difference of
relevant meson-antimeson mixing.
For example, contribution of  operator
$(\bar u \g_\mu c)(\bar \psi \g^\mu \g_5\psi)$
to width difference in $D^0$ - $\overline{D^0}$ is given by
$(\GG^{D^0})_{VA}$ given in Eq.~\eqref{eq: G12-fermion},
 which contains two terms related to different matrix elements:
 $\langle Q_S^{uc}\rangle$ and $\langle Q_V^{uc}\rangle$.
 In this case, the ``kink'' in the $y_{12}^D$ constraint
 corresponds to cancellation between contribution of these two matrix elements.

Finally, in Fig.~\ref{fig: fermion-DM-tensor} we show bounds on the tensor operators.
Again, as can be seen from the first and second column plots, $B$ meson decays provide far better constraints on these operators, and $\Delta\Gamma_{s, d}$
are only relevant for large $m_\psi$ values. But again in the charm sector, $y_{12}^D$ constraint is very important; we find that in the region $m_\psi \gtrsim 400$ MeV, the $y_{12}^D$ constraint is better than constraint from charm decays.

%%%%%%%%%%%%%%%%%%%%%%%%%%%%%%%%%%%%%%%%%%
%%%%%%%%%%%%%%%%%%%%%%%%%%%%%%%%%%%%%%%%%%
\section{Summary}\label{sec: summary}

Working in the EFT framework containing a light invisible particle
of spin $0$ or $1/2$ as an explicit degree of freedom,
we calculated the contributions of all relevant, effective operators to
the lifetime difference in beauty and charm mesons mixings. In our analysis,
we considered effective operators of dimension six
and focused on operators involving a pair of light NP particles.
We also pointed out that the previously considered effective operators containing 
a single dark sector particle do not contribute to the lifetime difference in neutral 
heavy mesons.

We compared the constraints resulting from the lifetime difference with
those from various beauty and charm meson decays. We find that in the beauty sector,
the bounds on the Wilson coefficients of relevant, effective operators obtained 
from the meson decays are typically more robust than those obtained from the 
lifetime difference. This implies that the contribution of the dark sector
particles to the lifetime difference in $B$ mesons is not large.

Contrary to that, for operators relevant to charm physics, we find that lifetime 
difference in charm mixing is crucial for constraining several operators when 
the NP particle is massive. We note that the lack of experimental data in charm 
decays, \emph{e.g.}, $D\to \rho\, \Emiss$, implies that vector-axial effective operators are 
presently not constrained. New constraints on these operators 
are obtained using the data on lifetime differences in charm mixing.

\section*{Acknowledgements}
This research was supported in part by the DOE grant DE-SC0024357. This work was 
performed in part at Aspen Center for Physics, which is supported by National 
Science Foundation grant PHY-2210452.

%%%%%%%%%%%%%%%%%%%%%%%%%%%%%%%%%%%%%%%%%%
%%%%%%%%%%%%%%%%%%%%%%%%%%%%%%%%%%%%%%%%%%
\appendix

%%%%%%%%%%%%%%%%%%%%%%%%%%%%%%%%%%%%%%%%%%
\section{\boldmath $\Delta F=2$ matrix elements}\label{app: DeltaF2}

In this Appendix we present some formulas relevant for computing the results
presented in Section \ref{sec: results} from the Eqs.~(\ref{eq: G12-scalar}) and 
(\ref{eq: G12-fermion}) of Section \ref{sec: lifetime-diff}. 

It is often useful to parametrize \cite{Petrov:2021idw} the four-fermion matrix 
elements relevant for the computation of the lifetime difference as
\begin{align}
	\langle {M^0} | (\bar q_i \g^\mu P_L Q_j)(\bar q_i \g_\mu P_L  Q_j)| \overline{M^0}\rangle &= \frac{2}{3}m^2_{M^0} f_{M^0}^2 B_{M^0}^{(1)}\,, 
\nonumber \\
	\langle {M^0} | (\bar q_i \g^\mu P_L Q_j)(\bar q_i \g_\mu P_R  Q_j)| \overline{M^0}\rangle &= -\frac{1}{3}\left[\left(\frac{m_{M^0}}{m_{q_i}(\mu) + m_{Q_j}(\mu)}\right)^2 + \frac{3}{2}\right] m^2_{M^0} f_{M^0}^2 B_{M^0}^{(5)}\,,
 \nonumber \\
	\langle {M^0} | (\bar q_i  P_L Q_j)(\bar q_i P_L  Q_j)| \overline{M^0}\rangle &= -\frac{5}{12}\left(\frac{m_{M^0}}{m_{q_i}(\mu) + m_{Q_j}(\mu)}\right)^2 m^2_{M^0} f_{M^0}^2 B_{M^0}^{(2)}\,,\\
		\langle {M^0} | (\bar q_i P_L Q_j)(\bar q_i P_R  Q_j)| \overline{M^0}\rangle &= \frac{1}{2}\left[\left(\frac{m_{M^0}}{m_{q_i}(\mu) + m_{Q_j}(\mu)}\right)^2 + \frac{1}{6}\right]  m^2_{M^0} f_{M^0}^2 B_{M^0}^{(4)}\,, 
\nonumber \\
	\langle {M^0} | (\bar q_i \sigma^{\mu\nu} P_L Q_j)(\bar q_i \sigma^{\mu\nu} P_L  Q_j)| \overline{M^0}\rangle &= \frac{1}{3}\left(\frac{m_{M^0}}{m_{q_i}(\mu) + m_{Q_j}(\mu)}\right)^2 m^2_{M^0} f_{M^0}^2 \left(5B_{M^0}^{(2)} - 2 B_{M^0}^{(3)}\right)\,,
\nonumber 
\end{align}
with meson states normalized as
$\langle M^0 | M^0\rangle = \langle \overline{M^0} | \overline{M^0}\rangle = 2 m_{M^0}$.
For evaluation of the matrix elements, we use \texttt{Flavio} package \cite{Straub:2018kue}, where
the numerical values for bag parameters 
$B_{M^0}^{(i)}$ for $B_s^0$ and $B^0$ are taken from Ref.~\cite{Greljo:2022jac} (based on results in Ref.~\cite{King:2019lal,Dowdall:2019bea}), and for $D^0$ meson from Ref.~\cite{Carrasco:2015pra}.
The values for decay constants $f_{M_0}$ are taken from
FLAG review \cite{FlavourLatticeAveragingGroupFLAG:2021npn} 
in four flavor scheme.

%%%%%%%%%%%%%%%%%%%%%%%%%
\section{\boldmath Decay width formulas}\label{app: decay-rates}

The differential decay rate of a heavy pseudoscalar meson $M$ decaying to a pseudoscalar ($P$) or
vector ($V$) meson plus a pair of scalar DM particle is given by \cite{He:2022ljo}
\begin{align}\label{eq: M2Pphiphi}
		\frac{d\Gamma (M \to P \phi\phi^\dagger)}{dq^2}
		&= \frac{(\kappa_\phi\,\lambda_{MP})^{1/2}}{256\pi^3 m_M^3}
			\left[\Delta_{MP}^2 f_0^2(q^2)\,|C_S^{q_iQ_j}|^2
			+ \frac{1}{3} \kappa_\phi\,\lambda_{MP}\,f_{+}^2(q^2)\, |C_V^{q_iQ_j}|^2\right]\,,\\
		\frac{d\Gamma (M \to V\phi\phi^\dagger)}{dq^2}
		&= \frac{(\kappa_\phi\,\lambda_{MV})^{1/2}}{256\pi^3 m_M^3}
			\left[\lambda_{MV} A_0^2(q^2)\,|C_P^{q_iQ_j}|^2
			+ \frac{2}{3}\frac{q^2 \kappa_\phi \,\lambda_{MV}}{(m_M +m_V )^2} V_{0}^2(q^2)\, |C_V^{q_iQ_j}|^2 \right.\nonumber\\
			&\hskip2.6cm +  \left.\frac{2}{3}\kappa_\phi \left\{(m_M+m_V)^2 q^2 A_1^2(q^2) + 32 m_M^2 m_V^2 A_{12}^2(q^2) |C_A^{q_iQ_j}|^2\right\}\right]\,,
\end{align}
where $\kappa_X = (1-4 \,m_X^2/q^2)$, $\Delta_{MM^\prime} = (m_{M}^2 - m_{M^\prime}^2)$,
and $\lambda_{MM^\prime}$ is shorthand notation for K\"all\'en function
$\lambda(m_M^2, m_{M^\prime}^2, q^2)\equiv m_M^4 + m_{M^\prime}^4 + q^4-2m_M^2m_{M^\prime}^2 -2 q^2m_{M^\prime}^2 - 2 q^2m_{M}^2$.
In writing the above expressions, we have ignored the mass of the
light quarks ($q_i=u, s$). In Eqs.~\eqref{eq: M2Pphiphi}-\eqref{eq: M2Vpsipsi}, 
functions $f_{0}, f_{+}, f_T$ and $V_0, A_0, A_1, A_2, A_{12}, T_1, T_2, T_{23}$ are
hadronic form factors parametrizing the matrix elements of  $M\to P$ and $M\to V$ transitions,
respectively. Their definitions can be found for example in Ref.~\cite{He:2022ljo}.
Specifically, we use form factors for $B \to K$ from Ref.~\cite{Gubernari:2023puw},
 $B\to\pi$ from Ref.~\cite{Leljak:2021vte}, $B \to K^\ast$ and $B\to \rho$
from Ref.~\cite{Bharucha:2015bzk}, and $D\to \pi$ from Ref.~\cite{Lubicz:2017syv}.
Note that if final state meson is  $\pi^0$ (or $\rho^0$), then, compared to $\pi^+$ (or $\rho^+$) case, decay rates in  Eqs.~\eqref{eq: M2Pphiphi}-\eqref{eq: M2Vpsipsi} have to be divided additionally by an overall factor of 2; this factor originates due to isospin symmetry relation between their form factors.  

Considering the fermionic DM scenario, the differential decay rate for $M\to P \psi \bar\psi$ is \cite{He:2023bnk}
\begin{align}\label{eq: M2Ppsipsi}
	&\frac{d\Gamma (M \to P \psi\bar\psi)}{dq^2}
		=  \frac{(\kappa_\psi\,\lambda_{MP})^{1/2}}{384\pi^3 m_M^3}\nonumber\\
			&\hskip0.5cm\times \left[\frac{3\,\Delta_{MP}^2 \,q^2}{(m_{Q_j} - m_{q_i})^2} f_0^2(q^2)
			\left\{\kappa_\psi|C_{SS}^{q_iQ_j}|^2 +   |C_{SP}^{q_iQ_j}|^2\right\}\right.  + \frac{2(q^2 + 2 m_\psi^2)\lambda_{MP}}{q^2} f_{+}^2(q^2) \,| C_{VV}^{q_iQ_j}|^2 \nonumber\\
			&\hskip01cm  
			 - \frac{12m_\psi \,\Delta_{MP}^2}{m_{Q_j} - m_{q_i}}f_{0}^2(q^2) \operatorname{Im}\left(C_{SP}^{q_iQ_j}C_{VA}^{q_iQ_j\ast}\right)
			 + \left\{\frac{12 m_\psi^2 \,\Delta_{MP}^2}{q^2} f_{0}^2(q^2) + 2 \kappa_\psi\lambda_{MP} f_{+}^2(q^2)\right\}|C_{VA}^{q_iQ_j}|^2\nonumber\\
			& \hskip1cm + \frac{4 \lambda_{MP}}{(m_M + m_P)^2}f_{T}^2(q^2)
			\left\{(q^2 + 8 m_\psi^2)|C_{TT}^{q_iQ_j}|^2 + q^2\kappa_\psi|C_{T\tilde{T}}^{q_iQ_j}|^2 \right\}\nonumber\\
			& \hskip1cm +\left. \frac{24 m_\psi \lambda_{MP}}{m_M + m_P}f_{+}(q^2)f_{T}(q^2)\operatorname{Re}\left(C_{VV}^{q_iQ_j}C_{TT}^{q_iQ_j\ast}\right)\right],
   \end{align}
and the corresponding expression for $M\to V \psi \bar\psi$ is (adapted from  Ref.~\cite{Bolton:2024egx})
   \begin{align}\label{eq: M2Vpsipsi}
    &\frac{d\Gamma (M \to V \psi\bar\psi)}{dq^2}
		=  \frac{(\kappa_\psi\,\lambda_{MV})^{1/2}}{96\pi^3 m_M^3}\nonumber\\
		& \hskip0.8cm\times 
			\left[\frac{\lambda_{MV}}{(M_M + m_V)^2}V_0^2(q^2)\left\{(q^2 + 2 m_\psi^2)|C_{VV}^{q_iQ_j}|^2 + q^2 \kappa_\psi|C_{VA}^{q_iQ_j}|^2\right\} \right.\nonumber\\
			& \hskip1.2cm + (q^2+2m_\psi^2)\left((m_M+m_V)^2 A_1^2(q^2) + \frac{32 m_M^2 m_V^2}{q^2} A_{12}^2(q^2)\right)|C_{AV}^{q_iQ_j}|^2\nonumber\\
			& \hskip1.2cm +\left\{q^2 \kappa_\psi\left((m_M+m_V)^2 A_1^2(q^2) + \frac{32m_M^2 m_V^2}{q^2}A_{12}^2(q^2)\right) + \frac{3m_\psi^2 \lambda_{MV}}{q^2}A_0^2(q^2)\right\}|C_{AA}^{q_iQ_j}|^2 \nonumber\\
			& \hskip1.2cm+ \frac{3}{4}\frac{\lambda_{MV}\, q^2}{(m_{Q_j}+m_{q_i})^2}A_0^2(q^2) \left\{ \kappa_\psi|C_{PS}^{q_iQ_j}|^2 +  |C_{PP}^{q_iQ_j}|^2\right\} \nonumber\\
			& \hskip1.2cm  +  \frac{2}{q^2}\left\{(q^2 + 8m_\psi^2) \lambda_{MV} T_1^2(q^2)
			+ q^2\kappa_\psi\left(\Delta_{MV}^2 T_2^2(q^2) + \frac{8m_M^2 m_V^2 q^2}{(m_M + m_V)^2}T_{23}^2(q^2)\right)\right\}|C_{TT}^{q_iQ_j}|^2\nonumber\\
			& \hskip1.2cm + \frac{2}{q^2}\left\{ q^2\kappa_\psi \lambda_{MV} T_1^2(q^2)
			+ (q^2+8m_\psi^2)\left(\Delta_{MV}^2 T_2^2(q^2) + \frac{8m_M^2 m_V^2 q^2}{(m_M + m_V)^2}T_{23}^2(q^2)\right)\right\}|C_{T\tilde{T}}^{q_iQ_j}|^2\nonumber\\
			& \hskip1.2cm - \frac{3 m_\psi\lambda_{MV}}{m_{Q_j}+m_{q_i}}A_0^2(q^2)    \operatorname{Re}\left(C_{AA}^{q_iQ_j}C_{PP}^{q_iQ_j\ast}\right)
			- \frac{12 m_\psi \lambda_{MV}}{m_M + m_V} V_0(q^2) T_1(q^2) \operatorname{Re}\left(C_{VV}^{q_iQ_j}C_{TT}^{q_iQ_j\ast}\right)\nonumber\\
			& \hskip1.2cm - \left. 12 m_\psi (m_M + m_V)\left\{\Delta_{MV}^2 A_1(q^2) T_2(q^2) + \frac{16 m_M^2 m_V^2}{(m_M + m_V)^2} A_{12}(q^2) T_{23}(q^2)\right\} \operatorname{Re}\left(C_{AV}^{q_iQ_j}C_{T\tilde{T}}^{q_iQ_j\ast}\right)\right].
\end{align}
Lastly, the invisible decay width of a pseudoscalar $M$ decaying to 
a pair of either scalar DM or fermionic DM particle, originally calculated in
Ref.~\cite{Badin:2010uh}, is given by \cite{Kamenik:2011vy}
\begin{align}
	\Gamma(M \to \phi \phi^\dagger) &= \frac{\beta_\phi f_{M}^2 m_{Q_j}^2 m_{M}^3}{16 \pi (m_{Q_j} + m_{q_i})^2}|C_P^{q_iQ_j}|^2,\\
	\Gamma(M \to \psi \bar\psi) &= \frac{\beta_\psi f_{M}^2}{8\pi}
			\left[\frac{ m_{M}^5}{(m_{Q_j} + m_{q_i})^2}\left(\beta_\psi^2|C_{PS}^{q_iQ_j}|^2 + |C_{PP}^{q_iQ_j}|^2\right) + 4m_\psi^2 m_{M} |C_{AA}^{q_iQ_j}|^2  \right.\nonumber\\
			  &\hskip2cm- \left.\frac{4 m_\psi m_{M}^3}{(m_{Q_j} + m_{q_i})} \operatorname{Re}\left(C_{AA}^{q_iQ_j}C_{PP}^{q_iQ_j\ast}\right)\right],
\end{align}
with $\beta_X$ defined below Eq.~\eqref{eq: G12-scalar}.

\bibliography{refs}{}

\providecommand{\href}[2]{#2}\begingroup\raggedright\begin{thebibliography}{10}

\bibitem{Planck:2015fie}
{\normalfont \bfseries Planck}, P.~A.~R. Ade {\it et al.}, ``{\it {Planck 2015
  results. XIII. Cosmological parameters}},''
  \href{http://dx.doi.org/10.1051/0004-6361/201525830}{Astron. Astrophys.
  {\normalfont \bfseries 594} (2016)  A13},
  \href{http://arxiv.org/abs/1502.01589}{{\normalfont \ttfamily
  arXiv:1502.01589}}.

\bibitem{Planck:2018vyg}
{\normalfont \bfseries Planck}, N.~Aghanim {\it et al.}, ``{\it {Planck 2018
  results. VI. Cosmological parameters}},''
  \href{http://dx.doi.org/10.1051/0004-6361/201833910}{Astron. Astrophys.
  {\normalfont \bfseries 641} (2020)  A6},
  \href{http://arxiv.org/abs/1807.06209}{{\normalfont \ttfamily
  arXiv:1807.06209}}. [Erratum: Astron.Astrophys. 652, C4 (2021)].

\bibitem{Clowe:2003tk}
D.~Clowe, A.~Gonzalez, and M.~Markevitch, ``{\it {Weak lensing mass
  reconstruction of the interacting cluster 1E0657-558: Direct evidence for the
  existence of dark matter}},''
  \href{http://dx.doi.org/10.1086/381970}{Astrophys. J. {\normalfont \bfseries
  604} (2004)  596--603},
  \href{http://arxiv.org/abs/astro-ph/0312273}{{\normalfont \ttfamily
  arXiv:astro-ph/0312273}}.

\bibitem{Rubin:1980zd}
V.~C. Rubin, N.~Thonnard, and W.~K. Ford, Jr., ``{\it {Rotational properties of
  21 SC galaxies with a large range of luminosities and radii, from NGC 4605
  (R=4kpc) to UGC 2885 (R=122kpc)}},''
  \href{http://dx.doi.org/10.1086/158003}{Astrophys. J. {\normalfont \bfseries
  238} (1980)  471}.

\bibitem{Bird:2004ts}
C.~Bird, P.~Jackson, R.~V. Kowalewski, and M.~Pospelov, ``{\it {Search for dark
  matter in $b \to s$ transitions with missing energy}},''
  \href{http://dx.doi.org/10.1103/PhysRevLett.93.201803}{Phys. Rev. Lett.
  {\normalfont \bfseries 93} (2004)  201803},
  \href{http://arxiv.org/abs/hep-ph/0401195}{{\normalfont \ttfamily
  arXiv:hep-ph/0401195}}.

\bibitem{Badin:2010uh}
A.~Badin and A.~A. Petrov, ``{\it {Searching for light Dark Matter in heavy
  meson decays}},'' \href{http://dx.doi.org/10.1103/PhysRevD.82.034005}{Phys.
  Rev. D {\normalfont \bfseries 82} (2010)  034005},
  \href{http://arxiv.org/abs/1005.1277}{{\normalfont \ttfamily
  arXiv:1005.1277}}.

\bibitem{Bhattacharya:2018msv}
B.~Bhattacharya, C.~M. Grant, and A.~A. Petrov, ``{\it {Invisible widths of
  heavy mesons}},'' \href{http://dx.doi.org/10.1103/PhysRevD.99.093010}{Phys.
  Rev. D {\normalfont \bfseries 99} (2019) no.~9, 093010},
  \href{http://arxiv.org/abs/1809.04606}{{\normalfont \ttfamily
  arXiv:1809.04606}}.

\bibitem{Altmannshofer:2009ma}
W.~Altmannshofer, A.~J. Buras, D.~M. Straub, and M.~Wick, ``{\it {New
  strategies for New Physics search in $B \to K^{*} \nu \bar{\nu}$, $B \to K
  \nu \bar{\nu}$ and $B \to X_{s} \nu \bar{\nu}$ decays}},''
  \href{http://dx.doi.org/10.1088/1126-6708/2009/04/022}{JHEP {\normalfont
  \bfseries 04} (2009)  022},
  \href{http://arxiv.org/abs/0902.0160}{{\normalfont \ttfamily
  arXiv:0902.0160}}.

\bibitem{Kamenik:2011vy}
J.~F. Kamenik and C.~Smith, ``{\it {FCNC portals to the dark sector}},''
  \href{http://dx.doi.org/10.1007/JHEP03(2012)090}{JHEP {\normalfont \bfseries
  03} (2012)  090}, \href{http://arxiv.org/abs/1111.6402}{{\normalfont
  \ttfamily arXiv:1111.6402}}.

\bibitem{Tandean:2019tkm}
J.~Tandean, ``{\it {Rare hyperon decays with missing energy}},''
  \href{http://dx.doi.org/10.1007/JHEP04(2019)104}{JHEP {\normalfont \bfseries
  04} (2019)  104}, \href{http://arxiv.org/abs/1901.10447}{{\normalfont
  \ttfamily arXiv:1901.10447}}.

\bibitem{Li:2020dpc}
G.~Li, T.~Wang, Y.~Jiang, J.-B. Zhang, and G.-L. Wang, ``{\it {Spin-$1/2$
  invisible particles in heavy meson decays}},''
  \href{http://dx.doi.org/10.1103/PhysRevD.102.095019}{Phys. Rev. D
  {\normalfont \bfseries 102} (2020) no.~9, 095019},
  \href{http://arxiv.org/abs/2004.10942}{{\normalfont \ttfamily
  arXiv:2004.10942}}.

\bibitem{Fajfer:2021woc}
S.~Fajfer and A.~Novosel, ``{\it {Colored scalars mediated rare charm meson
  decays to invisible fermions}},''
  \href{http://dx.doi.org/10.1103/PhysRevD.104.015014}{Phys. Rev. D
  {\normalfont \bfseries 104} (2021) no.~1, 015014},
  \href{http://arxiv.org/abs/2101.10712}{{\normalfont \ttfamily
  arXiv:2101.10712}}.

\bibitem{Li:2021sqe}
G.~Li, T.~Wang, J.-B. Zhang, and G.-L. Wang, ``{\it {The light invisible boson
  in FCNC decays of B and $B_c$ mesons}},''
  \href{http://dx.doi.org/10.1140/epjc/s10052-021-09333-z}{Eur. Phys. J. C
  {\normalfont \bfseries 81} (2021) no.~6, 564},
  \href{http://arxiv.org/abs/2103.12921}{{\normalfont \ttfamily
  arXiv:2103.12921}}.

\bibitem{Felkl:2021uxi}
T.~Felkl, S.~L. Li, and M.~A. Schmidt, ``{\it {A tale of invisibility:
  constraints on new physics in $b \to s\nu\nu$}},''
  \href{http://dx.doi.org/10.1007/JHEP12(2021)118}{JHEP {\normalfont \bfseries
  12} (2021)  118}, \href{http://arxiv.org/abs/2111.04327}{{\normalfont
  \ttfamily arXiv:2111.04327}}.

\bibitem{He:2022ljo}
X.-G. He, X.-D. Ma, and G.~Valencia, ``{\it {FCNC B and K meson decays with
  light bosonic Dark Matter}},''
  \href{http://dx.doi.org/10.1007/JHEP03(2023)037}{JHEP {\normalfont \bfseries
  03} (2023)  037}, \href{http://arxiv.org/abs/2209.05223}{{\normalfont
  \ttfamily arXiv:2209.05223}}.

\bibitem{Geng:2022kmf}
C.-Q. Geng and G.~Li, ``{\it {FCNC processes of charmed hadrons with invisible
  scalar}},'' \href{http://dx.doi.org/10.1016/j.physletb.2023.137811}{Phys.
  Lett. B {\normalfont \bfseries 839} (2023)  137811},
  \href{http://arxiv.org/abs/2212.04699}{{\normalfont \ttfamily
  arXiv:2212.04699}}.

\bibitem{Li:2023sjf}
G.~Li and J.~Tandean, ``{\it {FCNC charmed-hadron decays with invisible singlet
  particles in light of recent data}},''
  \href{http://dx.doi.org/10.1007/JHEP11(2023)205}{JHEP {\normalfont \bfseries
  11} (2023)  205}, \href{http://arxiv.org/abs/2306.05333}{{\normalfont
  \ttfamily arXiv:2306.05333}}.

\bibitem{Felkl:2023ayn}
T.~Felkl, A.~Giri, R.~Mohanta, and M.~A. Schmidt, ``{\it {When energy goes
  missing: new physics in $b\to s \nu \nu$ with sterile neutrinos}},''
  \href{http://dx.doi.org/10.1140/epjc/s10052-023-12326-9}{Eur. Phys. J. C
  {\normalfont \bfseries 83} (2023) no.~12, 1135},
  \href{http://arxiv.org/abs/2309.02940}{{\normalfont \ttfamily
  arXiv:2309.02940}}.

\bibitem{He:2023bnk}
X.-G. He, X.-D. Ma, and G.~Valencia, ``{\it {Revisiting models that enhance
  $B^{+}\to K^{+} \nu\nu$ in light of the new Belle II measurement}},''
  \href{http://dx.doi.org/10.1103/PhysRevD.109.075019}{Phys. Rev. D
  {\normalfont \bfseries 109} (2024) no.~7, 075019},
  \href{http://arxiv.org/abs/2309.12741}{{\normalfont \ttfamily
  arXiv:2309.12741}}.

\bibitem{Gabrielli:2024wys}
E.~Gabrielli, L.~Marzola, K.~M\"u\"ursepp, and M.~Raidal, ``{\it {Explaining
  the $B^+\rightarrow K^+ \nu \bar{\nu }$ excess via a massless dark
  photon}},'' \href{http://dx.doi.org/10.1140/epjc/s10052-024-12818-2}{Eur.
  Phys. J. C {\normalfont \bfseries 84} (2024) no.~5, 460},
  \href{http://arxiv.org/abs/2402.05901}{{\normalfont \ttfamily
  arXiv:2402.05901}}.

\bibitem{Bolton:2024egx}
P.~D. Bolton, S.~Fajfer, J.~F. Kamenik, and M.~Novoa-Brunet, ``{\it {Signatures
  of Light New Particles in $B\to K^{(*)} E_{\rm miss}$}},''
  \href{http://arxiv.org/abs/2403.13887}{{\normalfont \ttfamily
  arXiv:2403.13887}}.

\bibitem{Buras:2024ewl}
A.~J. Buras, J.~Harz, and M.~A. Mojahed, ``{\it {Disentangling new physics in
  $K\rightarrow\pi\bar{\nu}\nu$ and $B\rightarrow K(K^*)\bar{\nu}\nu$
  observables}},'' \href{http://arxiv.org/abs/2405.06742}{{\normalfont
  \ttfamily arXiv:2405.06742}}.

\bibitem{Belle:2016qek}
{\normalfont \bfseries Belle}, Y.~T. Lai {\it et al.}, ``{\it {Search for
  $D^{0}$ decays to invisible final states at Belle}},''
  \href{http://dx.doi.org/10.1103/PhysRevD.95.011102}{Phys. Rev. D {\normalfont
  \bfseries 95} (2017) no.~1, 011102},
  \href{http://arxiv.org/abs/1611.09455}{{\normalfont \ttfamily
  arXiv:1611.09455}}.

\bibitem{Burdman:2001tf}
G.~Burdman, E.~Golowich, J.~L. Hewett, and S.~Pakvasa, ``{\it {Rare charm
  decays in the standard model and beyond}},''
  \href{http://dx.doi.org/10.1103/PhysRevD.66.014009}{Phys. Rev. D {\normalfont
  \bfseries 66} (2002)  014009},
  \href{http://arxiv.org/abs/hep-ph/0112235}{{\normalfont \ttfamily
  arXiv:hep-ph/0112235}}.

\bibitem{BESIII:2021slf}
{\normalfont \bfseries BESIII}, M.~Ablikim {\it et al.}, ``{\it {Search for the
  decay $D^{0} \to \pi^{0} \nu \bar{\nu}$}},''
  \href{http://dx.doi.org/10.1103/PhysRevD.105.L071102}{Phys. Rev. D
  {\normalfont \bfseries 105} (2022) no.~7, L071102},
  \href{http://arxiv.org/abs/2112.14236}{{\normalfont \ttfamily
  arXiv:2112.14236}}.

\bibitem{Alonso-Alvarez:2023mgc}
G.~Alonso-\'Alvarez and M.~Escudero~Abenza, ``{\it {The first limit on
  invisible decays of $B_s$ mesons comes from LEP}},''
  \href{http://dx.doi.org/10.1140/epjc/s10052-024-12936-x}{Eur. Phys. J. C
  {\normalfont \bfseries 84} (2024) no.~5, 553},
  \href{http://arxiv.org/abs/2310.13043}{{\normalfont \ttfamily
  arXiv:2310.13043}}.

\bibitem{Belle-II:2023esi}
{\normalfont \bfseries Belle-II}, I.~Adachi {\it et al.}, ``{\it {Evidence for
  $B^{+}\to K^{+}\nu\bar{\nu}$ Decays}},''
  \href{http://arxiv.org/abs/2311.14647}{{\normalfont \ttfamily
  arXiv:2311.14647}}.

\bibitem{Belle:2017oht}
{\normalfont \bfseries Belle}, J.~Grygier {\it et al.}, ``{\it {Search for
  $B\to h\nu\bar{\nu}$ decays with semileptonic tagging at Belle}},''
  \href{http://dx.doi.org/10.1103/PhysRevD.96.091101}{Phys. Rev. D {\normalfont
  \bfseries 96} (2017) no.~9, 091101},
  \href{http://arxiv.org/abs/1702.03224}{{\normalfont \ttfamily
  arXiv:1702.03224}}. [Addendum: Phys.Rev.D 97, 099902 (2018)].

\bibitem{Belle:2013tnz}
{\normalfont \bfseries Belle}, O.~Lutz {\it et al.}, ``{\it {Search for $B \to
  h^{(*)} \nu \bar{\nu}$ with the full Belle $\Upsilon(4S)$ data sample}},''
  \href{http://dx.doi.org/10.1103/PhysRevD.87.111103}{Phys. Rev. D {\normalfont
  \bfseries 87} (2013) no.~11, 111103},
  \href{http://arxiv.org/abs/1303.3719}{{\normalfont \ttfamily
  arXiv:1303.3719}}.

\bibitem{BaBar:2012yut}
{\normalfont \bfseries BaBar}, J.~P. Lees {\it et al.}, ``{\it {Improved Limits
  on $B^0$ Decays to Invisible Final States and to $\nu \bar{\nu} \gamma$}},''
  \href{http://dx.doi.org/10.1103/PhysRevD.86.051105}{Phys. Rev. D {\normalfont
  \bfseries 86} (2012)  051105},
  \href{http://arxiv.org/abs/1206.2543}{{\normalfont \ttfamily
  arXiv:1206.2543}}.

\bibitem{Straub:2018kue}
D.~M. Straub, ``{\it {flavio: a Python package for flavour and precision
  phenomenology in the Standard Model and beyond}},''
  \href{http://arxiv.org/abs/1810.08132}{{\normalfont \ttfamily
  arXiv:1810.08132}}.

\bibitem{ALEPH:2000vvi}
{\normalfont \bfseries ALEPH}, R.~Barate {\it et al.}, ``{\it {Measurements of
  BR$(b\to \tau^{-} \bar{\nu}_\tau X)$ and BR$(b \to \tau^{-} \bar{\nu}_\tau
  D^{\ast \pm} X)$ and upper limits on BR$(B^{-} \to \tau^{-} \bar{\nu}_\tau)$
  and BR$(b\to s \nu \bar{\nu})$}},''
  \href{http://dx.doi.org/10.1007/s100520100612}{Eur. Phys. J. C {\normalfont
  \bfseries 19} (2001)  213--227},
  \href{http://arxiv.org/abs/hep-ex/0010022}{{\normalfont \ttfamily
  arXiv:hep-ex/0010022}}.

\bibitem{Golowich:2006gq}
E.~Golowich, S.~Pakvasa, and A.~A. Petrov, ``{\it {New Physics contributions to
  the lifetime difference in $D^0$-$\overline{D}^0$ mixing}},''
  \href{http://dx.doi.org/10.1103/PhysRevLett.98.181801}{Phys. Rev. Lett.
  {\normalfont \bfseries 98} (2007)  181801},
  \href{http://arxiv.org/abs/hep-ph/0610039}{{\normalfont \ttfamily
  arXiv:hep-ph/0610039}}.

\bibitem{Golowich:2007ka}
E.~Golowich, J.~Hewett, S.~Pakvasa, and A.~A. Petrov, ``{\it {Implications of
  $D^0$-$\bar{D}^0$ Mixing for New Physics}},''
  \href{http://dx.doi.org/10.1103/PhysRevD.76.095009}{Phys. Rev. D {\normalfont
  \bfseries 76} (2007)  095009},
  \href{http://arxiv.org/abs/0705.3650}{{\normalfont \ttfamily
  arXiv:0705.3650}}.

\bibitem{Golowich:1998pz}
E.~Golowich and A.~A. Petrov, ``{\it {Can nearby resonances enhance D0 -
  anti-D0 mixing?}},''
  \href{http://dx.doi.org/10.1016/S0370-2693(98)00329-3}{Phys. Lett. B
  {\normalfont \bfseries 427} (1998)  172--178},
  \href{http://arxiv.org/abs/hep-ph/9802291}{{\normalfont \ttfamily
  arXiv:hep-ph/9802291}}.

\bibitem{Lehmann:2020lcv}
B.~V. Lehmann and S.~Profumo, ``{\it {Cosmology and prospects for sub-MeV dark
  matter in electron recoil experiments}},''
  \href{http://dx.doi.org/10.1103/PhysRevD.102.023038}{Phys. Rev. D
  {\normalfont \bfseries 102} (2020) no.~2, 023038},
  \href{http://arxiv.org/abs/2002.07809}{{\normalfont \ttfamily
  arXiv:2002.07809}}.

\bibitem{Cutkosky:1960sp}
R.~E. Cutkosky, ``{\it {Singularities and discontinuities of Feynman
  amplitudes}},'' \href{http://dx.doi.org/10.1063/1.1703676}{J. Math. Phys.
  {\normalfont \bfseries 1} (1960)  429--433}.

\bibitem{Patel:2016fam}
H.~H. Patel, ``{\it {Package-X 2.0: A Mathematica package for the analytic
  calculation of one-loop integrals}},''
  \href{http://dx.doi.org/10.1016/j.cpc.2017.04.015}{Comput. Phys. Commun.
  {\normalfont \bfseries 218} (2017)  66--70},
  \href{http://arxiv.org/abs/1612.00009}{{\normalfont \ttfamily
  arXiv:1612.00009}}.

\bibitem{Artuso:2015swg}
M.~Artuso, G.~Borissov, and A.~Lenz, ``{\it {CP violation in the $B_s^0$
  system}},'' \href{http://dx.doi.org/10.1103/RevModPhys.88.045002}{Rev. Mod.
  Phys. {\normalfont \bfseries 88} (2016) no.~4, 045002},
  \href{http://arxiv.org/abs/1511.09466}{{\normalfont \ttfamily
  arXiv:1511.09466}}. [Addendum: Rev.Mod.Phys. 91, 049901 (2019)].

\bibitem{Beneke:2003az}
M.~Beneke, G.~Buchalla, A.~Lenz, and U.~Nierste, ``{\it {CP asymmetry in flavor
  specific B decays beyond leading logarithms}},''
  \href{http://dx.doi.org/10.1016/j.physletb.2003.09.089}{Phys. Lett. B
  {\normalfont \bfseries 576} (2003)  173--183},
  \href{http://arxiv.org/abs/hep-ph/0307344}{{\normalfont \ttfamily
  arXiv:hep-ph/0307344}}.

\bibitem{Albrecht:2024oyn}
J.~Albrecht, F.~Bernlochner, A.~Lenz, and A.~Rusov, ``{\it {Lifetimes of
  b-hadrons and mixing of neutral B-mesons: theoretical and experimental
  status}},'' \href{http://dx.doi.org/10.1140/epjs/s11734-024-01124-3}{Eur.
  Phys. J. ST {\normalfont \bfseries 233} (2024) no.~2, 359--390},
  \href{http://arxiv.org/abs/2402.04224}{{\normalfont \ttfamily
  arXiv:2402.04224}}.

\bibitem{ParticleDataGroup:2022pth}
{\normalfont \bfseries Particle Data Group}, R.~L. Workman {\it et al.}, ``{\it
  {Review of Particle Physics}},''
  \href{http://dx.doi.org/10.1093/ptep/ptac097}{PTEP {\normalfont \bfseries
  2022} (2022)  083C01}.

\bibitem{Inami:1980fz}
T.~Inami and C.~S. Lim, ``{\it {Effects of Superheavy Quarks and Leptons in
  Low-Energy Weak Processes $K_L \to \mu\bar\mu$, $K^{+} \to \pi^{+}
  \nu\bar\nu$ and $K^0 \leftrightarrow \overline{K^0}$}},''
  \href{http://dx.doi.org/10.1143/PTP.65.297}{Prog. Theor. Phys. {\normalfont
  \bfseries 65} (1981)  297}. [Erratum: Prog.Theor.Phys. 65, 1772 (1981)].

\bibitem{Buras:1990fn}
A.~J. Buras, M.~Jamin, and P.~H. Weisz, ``{\it {Leading and Next-to-leading
  {QCD} Corrections to $\epsilon$ Parameter and $B^0 - \bar{B}^0$ Mixing in the
  Presence of a Heavy Top Quark}},''
  \href{http://dx.doi.org/10.1016/0550-3213(90)90373-L}{Nucl. Phys. B
  {\normalfont \bfseries 347} (1990)  491--536}.

\bibitem{HFLAV:2022esi}
{\normalfont \bfseries HFLAV}, Y.~S. Amhis {\it et al.}, ``{\it {Averages of
  b-hadron, c-hadron, and $\tau$-lepton properties as of 2021}},''
  \href{http://dx.doi.org/10.1103/PhysRevD.107.052008}{Phys. Rev. D
  {\normalfont \bfseries 107} (2023) no.~5, 052008},
  \href{http://arxiv.org/abs/2206.07501}{{\normalfont \ttfamily
  arXiv:2206.07501}}. (online update is available at
  \url{https://hflav.web.cern.ch/}).

\bibitem{Lenz:2020awd}
A.~Lenz and G.~Wilkinson, ``{\it {Mixing and CP Violation in the Charm
  System}},'' \href{http://dx.doi.org/10.1146/annurev-nucl-102419-124613}{Ann.
  Rev. Nucl. Part. Sci. {\normalfont \bfseries 71} (2021)  59--85},
  \href{http://arxiv.org/abs/2011.04443}{{\normalfont \ttfamily
  arXiv:2011.04443}}.

\bibitem{Petrov:2021idw}
A.~A. Petrov, \href{http://dx.doi.org/10.1201/9781351176019}{{\it {Indirect
  Searches for New Physics}}}.
\newblock CRC Press, Boca Raton, 5, 2021.

\bibitem{Greljo:2022jac}
A.~Greljo, J.~Salko, A.~Smolkovi\v{c}, and P.~Stangl, ``{\it {Rare b decays
  meet high-mass Drell-Yan}},''
  \href{http://dx.doi.org/10.1007/JHEP05(2023)087}{JHEP {\normalfont \bfseries
  05} (2023)  087}, \href{http://arxiv.org/abs/2212.10497}{{\normalfont
  \ttfamily arXiv:2212.10497}}.

\bibitem{King:2019lal}
D.~King, A.~Lenz, and T.~Rauh, ``{\it {B$_{s}$ mixing observables and
  |V$_{td}$/V$_{ts}$| from sum rules}},''
  \href{http://dx.doi.org/10.1007/JHEP05(2019)034}{JHEP {\normalfont \bfseries
  05} (2019)  034}, \href{http://arxiv.org/abs/1904.00940}{{\normalfont
  \ttfamily arXiv:1904.00940}}.

\bibitem{Dowdall:2019bea}
R.~J. Dowdall, C.~T.~H. Davies, R.~R. Horgan, G.~P. Lepage, C.~J. Monahan,
  J.~Shigemitsu, and M.~Wingate, ``{\it {Neutral B-meson mixing from full
  lattice QCD at the physical point}},''
  \href{http://dx.doi.org/10.1103/PhysRevD.100.094508}{Phys. Rev. D
  {\normalfont \bfseries 100} (2019) no.~9, 094508},
  \href{http://arxiv.org/abs/1907.01025}{{\normalfont \ttfamily
  arXiv:1907.01025}}.

\bibitem{Carrasco:2015pra}
{\normalfont \bfseries ETM}, N.~Carrasco, P.~Dimopoulos, R.~Frezzotti,
  V.~Lubicz, G.~C. Rossi, S.~Simula, and C.~Tarantino, ``{\it
  {\ensuremath{\Delta}S=2 and \ensuremath{\Delta}C=2 bag parameters in the
  standard model and beyond from N$_f$=2+1+1 twisted-mass lattice QCD}},''
  \href{http://dx.doi.org/10.1103/PhysRevD.92.034516}{Phys. Rev. D {\normalfont
  \bfseries 92} (2015) no.~3, 034516},
  \href{http://arxiv.org/abs/1505.06639}{{\normalfont \ttfamily
  arXiv:1505.06639}}.

\bibitem{FlavourLatticeAveragingGroupFLAG:2021npn}
{\normalfont \bfseries Flavour Lattice Averaging Group (FLAG)}, Y.~Aoki {\it et
  al.}, ``{\it {FLAG Review 2021}},''
  \href{http://dx.doi.org/10.1140/epjc/s10052-022-10536-1}{Eur. Phys. J. C
  {\normalfont \bfseries 82} (2022) no.~10, 869},
  \href{http://arxiv.org/abs/2111.09849}{{\normalfont \ttfamily
  arXiv:2111.09849}}.

\bibitem{Gubernari:2023puw}
N.~Gubernari, M.~Reboud, D.~van Dyk, and J.~Virto, ``{\it {Dispersive analysis
  of $B \to K^{(*)}$ and $B_{s} \to \phi$ form factors}},''
  \href{http://dx.doi.org/10.1007/JHEP12(2023)153}{JHEP {\normalfont \bfseries
  12} (2023)  153}, \href{http://arxiv.org/abs/2305.06301}{{\normalfont
  \ttfamily arXiv:2305.06301}}.

\bibitem{Leljak:2021vte}
D.~Leljak, B.~Meli\'c, and D.~van Dyk, ``{\it {The $\overline{B}\to \pi$ form
  factors from QCD and their impact on $|V_{ub}|$}},''
  \href{http://dx.doi.org/10.1007/JHEP07(2021)036}{JHEP {\normalfont \bfseries
  07} (2021)  036}, \href{http://arxiv.org/abs/2102.07233}{{\normalfont
  \ttfamily arXiv:2102.07233}}.

\bibitem{Bharucha:2015bzk}
A.~Bharucha, D.~M. Straub, and R.~Zwicky, ``{\it {$B\to V\ell^+\ell^-$ in the
  Standard Model from light-cone sum rules}},''
  \href{http://dx.doi.org/10.1007/JHEP08(2016)098}{JHEP {\normalfont \bfseries
  08} (2016)  098}, \href{http://arxiv.org/abs/1503.05534}{{\normalfont
  \ttfamily arXiv:1503.05534}}.

\bibitem{Lubicz:2017syv}
{\normalfont \bfseries ETM}, V.~Lubicz, L.~Riggio, G.~Salerno, S.~Simula, and
  C.~Tarantino, ``{\it {Scalar and vector form factors of $D \to \pi(K) \ell
  \nu$ decays with $N_f=2+1+1$ twisted fermions}},''
  \href{http://dx.doi.org/10.1103/PhysRevD.96.054514}{Phys. Rev. D {\normalfont
  \bfseries 96} (2017) no.~5, 054514},
  \href{http://arxiv.org/abs/1706.03017}{{\normalfont \ttfamily
  arXiv:1706.03017}}. [Erratum: Phys.Rev.D 99, 099902 (2019), Erratum:
  Phys.Rev.D 100, 079901 (2019)].

\end{thebibliography}\endgroup
\bibliographystyle{utcaps_mod}

\end{document}